\documentclass[sigconf,screen]{acmart}

\usepackage{amsmath}
\usepackage[ruled,linesnumbered]{algorithm2e}
\usepackage{listings}
\usepackage{subcaption}
\usepackage{mwe}
\usepackage{multicol}

\graphicspath{ {./images/} }

\AtBeginDocument{%
  \providecommand\BibTeX{{%
    \normalfont B\kern-0.5em{\scshape i\kern-0.25em b}\kern-0.8em\TeX}}}

\setcopyright{acmcopyright}
\copyrightyear{2022}
\acmYear{2022}
\acmISBN{978-1-4503-9629-5}
\acmPrice{}
\acmDOI{}

\acmConference[HP3C]{Inproceedings}{Jilin}{China}

\begin{document}

\title{Modern Distributed Data-Parallel Large-Scale Pre-training Strategies For NLP models}

\author{Hao Bai}
\email{haob2@illinois.edu}
\orcid{0000-0001-9723-7490}
\affiliation{
  \institution{University of Illinois Urbana-Champaign}
  \streetaddress{506 S. Wright St.}
  \city{Champaign}
  \state{Illinois}
  \country{USA}
  \postcode{61801}
}

\renewcommand{\shortauthors}{Bai}

\begin{abstract}

	Distributed deep learning is becoming increasingly popular due to the expanding demand for computing resources for deep learning models with a larger amount of parameters. Different from traditional training approaches, data-parallel training allows multiple compute nodes to train large deep learning models simultaneously in order to boost the training efficiency. In this paper, we present and compare six strategies for data-parallel training using PyTorch on the language model GPT-2 with 100M parameters using a qualitative approach. These strategies are Single GPU, Single Parameter Server, Distributed Parameter Server, Horovod, Distributed Parameter Server with Apex mixed-precision strategy, and Horovod with Apex mixed-precision strategy. We also analyze the quantitative experiment results from each strategy. In the end, we draw the conclusion that the Distributed Parameter Server with Apex mixed-precision strategy has the best performance on single node training, while Horovod with Apex is the most robust approach to use when we have single or multiple nodes.
	
\end{abstract}




\maketitle

\section{Introduction} \label{sec:intro}

Distributed deep learning is continuously gaining popularity due to its performance improvements on very-large-scale (VLS) deep learning models, including Computer Vision and Natural Language Processing \cite{chahal2020hitchhiker}. This paper analyzes the state-of-the-art (SOTA) distributed deep learning strategies using conceptual abstractions and underlying mathematical principles, aiming to help researchers and engineers implement the strategies and accelerate the training process. 

Since the proposal of the distributed deep learning idea, the two largest deep learning framework, PyTorch \cite{paszke2019pytorch} and TensorFlow \cite{abadi2016tensorflow}, have been working on practical ways to implement the theories concluded by researches proposing this idea \cite{abadi2016tensorflowlarge, li2020pytorch}, with the goal to minimize the changes to the code already written. 

Research has been done to test the experimental results of the practical performance of the theories. However, research concerning experimental results and a comprehensive comparison using the most recent tools by PyTorch using NVIDIA Graphical Processing Units (GPUs) is rare. This forces researchers to guess when choosing which strategy to use, especially for those who tackle with VLS models. In our research, we implemented the six most popular strategies in PyTorch to utilize data parallelization for distributed large-scale deep learning using GPT-2 model \cite{Radford2019}, and showed the first-hand experimental results of all strategies with reasoning and analysis.

Modern distributed VLS deep learning is divided mainly into two categories: data parallelization and model parallelization \cite{dean2012large}. As implied by their names, data parallelization is used to distribute data on different devices for parallelization, while model parallelization is to distribute parts of the model on different devices. In other words, they both apportion parts of information to multiple devices.

Each type of parallelization has its own intent. Data parallelization can accelerate the training process and increase the batch size to an approximately linear extent because the data is apportioned onto different devices, while model parallelization is invented to reduce the memory usage on one device. The usage and mathematical insight of them can be found in Appendix \ref{sec:data-parallel} and \ref{sec:model-parallel}. In our research, we mainly focus on data parallelization, with model parallelization to be addressed in later research.

\section{Related Work} \label{sec:related-works}

In this part, we present the prerequisites for understanding our work. First, we explain why we selected the six data-parallel strategies by surveying the SOTA distributed frameworks and their support for distributed deep learning. Second, as this research is conducted on the \textit{HAL} mini-super-computer at the National Center for Supercomputing Applications (NCSA) \cite{kindratenko2020hal}, the experimental results are highly related to the architecture and performance of this machine, which we're going to analyze in Section \ref{sec:data}. We also analyze the language model we're using in this study.

\subsection{State-of-the-art Data Parallel Frameworks} \label{sec:sota}

Here we provide a brief survey on the SOTA data-parallel frameworks and how they support PyTorch and Tensorflow. The frameworks include PyTorch, Tensorflow, Spark, Horovod, Ray, Hadoop, and Kubernetes, as shown in Table \ref{tab:sotas}.\footnote{We don't discuss the details of the algorithms of each framework, because this paper puts most efforts into practical strategies instead of algorithms.}

\begin{center}
\begin{table}[htp!]
\centering
\small
\setlength\tabcolsep{2.3pt}
\begin{tabular}{c|c|c}
\hline
&\multicolumn{2}{c}{Distribution Support}\\
\hline
Framework & PyTorch & Tensorflow \\
\hline
PyTorch* & Official & None \\
Tensorflow* & None & Official \\
Horovod* & Official & Official \\
Ray & Official & Official \\
\hline
Kubernetes & TorchElastic & Kubeflow \\
Spark & SparkTorch & TensorFlow on Spark \\
Hadoop & None & TensorFlow on Yarn \\
\hline
\end{tabular}
\caption{Data-parallel strategies on SOTA distributed systems and their relative complexity. "*": frameworks designed for deep learning. "None": no reliable support.}
\label{tab:sotas}
\end{table}
\end{center}

Among all the SOTA frameworks, PyTorch and Tensorflow are designed to be deep learning frameworks, and all of them offer original support for distributed deep learning tasks. Horovod is specifically designed to enable distributed deep leanirng using PyTorch or Tensorflow. Ray is a framework to scale compute-intensive workloads, but still officially supports distributed PyTorch and Tensorflow. All the three frameworks mentioned above offer reliable official support to distributed PyTorch and Tensorflow.

Kubernetes is a framework mostly for managing containerized applications, but also supports distributed deep learning with successful projects like \textit{Kubeflow} and \textit{TorchElastic}. Spark and Hadoop are two frameworks proposed based on the MapReduce algorithm and have the least to do with deep learning, so their support for PyTorch and Tensorflow is even worse. 

\subsection{High-Performance Computing and \textit{HAL}} \label{sec:hal}

High-Performance Computing (HPC) relies on supercomputers to solve computationally demanding tasks. In the terminology of HPC, a computer with multiple CPU cores and GPUs is called a \textit{worker node} and a group of these computers is called a \textit{cluster}, which will be frequently used in this paper. HPC is crucial in distributed deep learning because we need a strong hardware backbone for multi-worker training. 

HPC systems have multiple computing nodes as they are the essential part of computing resources. For example, \textit{HAL} is such a mini supercomputer consisting of shared storage, one login node, 16 computing nodes, each consisting of four GPUs and two 20-core CPUs. This research is conducted using \textit{HAL}.
		
\textit{HAL} is constructed using PowerPC CPUs and Tesla V100 GPUs, and it uses SLURM to manage the task queue.\footnote{More information about SLURM is listed in Appendix \ref{sec:slurm}.} \textit{HAL} uses \texttt{conda} instead of \texttt{pip} for package management. \texttt{Pip} is still available and we can use it as well, typically when the package is not available through \texttt{conda} and local compilation\footnote{Local compiling means downloading the source code from the official site, then compiling and install it locally.}.
	
\subsection{The GPT Language Models} \label{sec:gpt} 

Proposed by OpenAI, the Generative Pre-trained Transformer (GPT) models have been very successful in Natural Language Generation (NLG) tasks \cite{Radford2019, Brown2020}. Different from traditional language models, the GPT language models are designed to be stackable - the model can be constructed by stacking multi-head decoder layers proposed by the transformer architecture. 

The transformer is one of the most famous encoder-decoder architectures proposed by Google, which aims to train a very deep model using attention layers. Both encoder and decoder in the transformer architecture are stacks of multi-head attention layers, as illustrated in Figure \ref{fig:gpt-2}. Some insights into the multi-head attention layer are elaborated in Appendix \ref{sec:transformer}. 

\begin{figure}[htp!]
	\begin{center}
  		\includegraphics[width=0.85\linewidth]{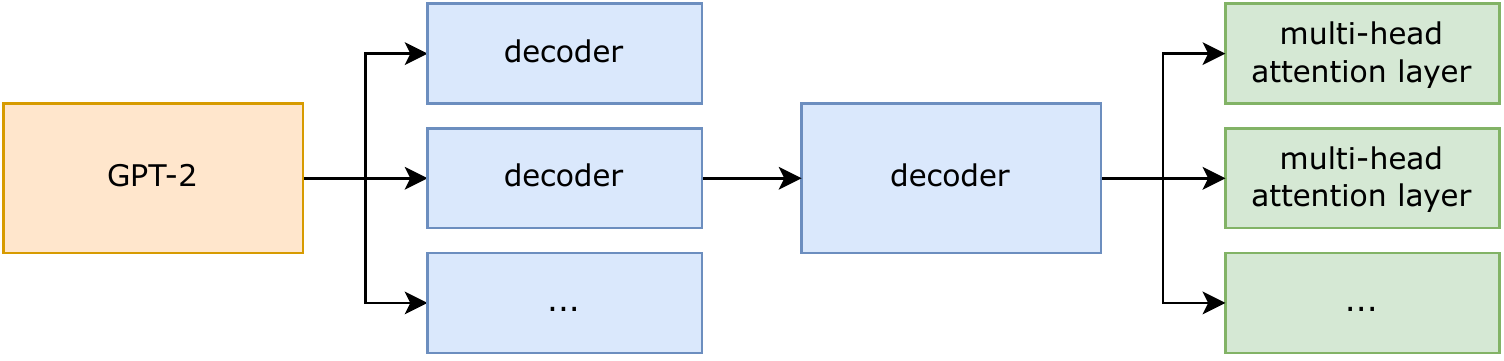}
	\end{center}
   	\caption{The conceptual architecture of the GPT-2 model is a stack of decoders, with each decoder stacked by multi-head attention layers.}
	\label{fig:gpt-2}
\end{figure}

We select GPT-2 as our subject for data-parallel training, because GPT-2 is well-supported by the PyTorch community and it is stackable, which offers great convenience for fitting into the device.\footnote{Stackability enables the model to be enlarged or shrunk without changing the model input/output, but simply hyper-parameters.} The scale of GPT-2 is also suitable for data-parallel training. Compared to GPT-2, although the GPT-3 model was proved to be a great success, the underlying model of GPT-3 was never released, and the scale of GPT-3 is too large to distribute for a mini-super-computer.

\section{Data Parallelization Variances} \label{sec:data}

As researchers have found that deep learning models with more parameters can store more information and improve model performance \cite{Dean2015}, models with more and more parameters are being proposed, especially for language models \cite{Roberts2020}. For example, the Google T5 language model reached 220 million parameters in 2019 \cite{Raffel2019}, the GPT-3 model consisted of 175 billion parameters in 2020 \cite{Brown2020}, and the Google Switch Transformer language model rose up to 1.6 trillion parameters in 2021. Every year, the sizes of the language models are expanding at a thousand-fold speed, leading to an urgent upgrade of the distributed systems to train them.

As mentioned in Section \ref{sec:intro}, the data-parallel strategies are used to increase the training throughput and shorten the time needed for training. However, the workers (GPUs) need a communication interface to synchronize the training data and updated model parameters, which leads to the introduction of the parameter server architecture. 

\subsection{Parameter Server Architectures} \label{sec:ps-archi}

\begin{figure}[htp!]
	\begin{center}
  		\includegraphics[width=0.9\linewidth]{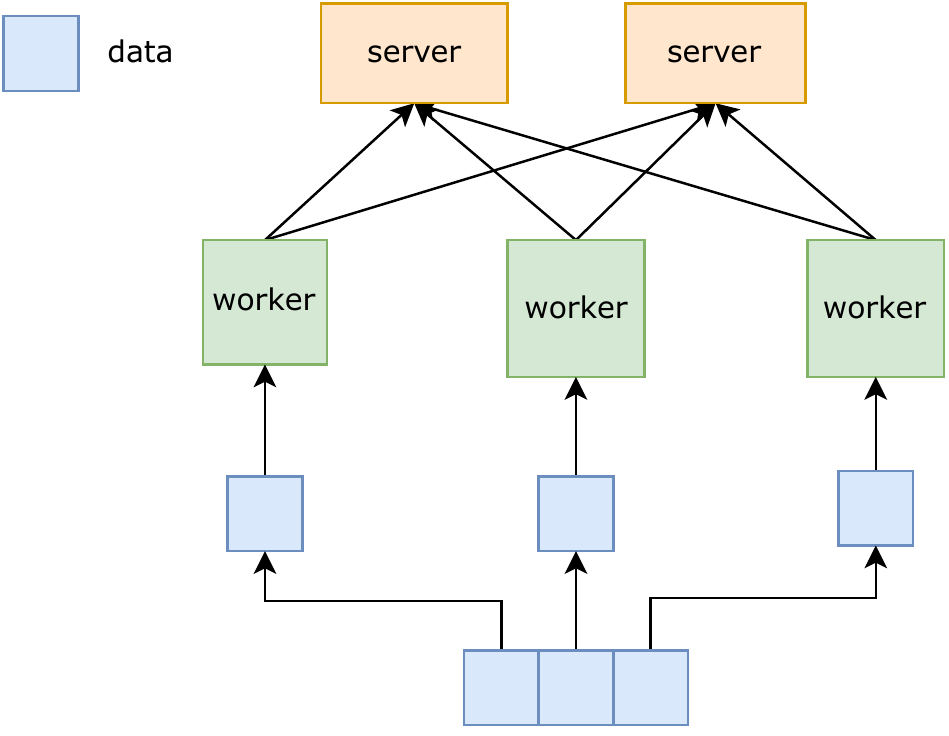}
	\end{center}
   	\caption{Abstract model of the parameter server architecture.}
	\label{fig:PS}
\end{figure}

The parameter server architecture is composed of the parameter server nodes and the worker (client) nodes \cite{Li2013}. The clients are in charge of learning the parameters according to the data scattered to them and updating the model parameters using the learning results, while the parameter servers store the updated parameters and manage synchronization. The abstract model of the parameter server architecture is shown in Figure \ref{fig:PS}.

According to the number of parameter servers, the data-parallel strategies are split into two main categories: Single Parameter Server (SPS) and Distributed Parameter Server (DPS) strategies. The SPS strategy utilizes one centralized server to authorize the synchronized parameters, while the DPS strategy sets up multiple parameter servers with well-defined typologies to minimize data transmission overhead. In the context of this paper, we denote DPS as a fully distributed one, i.e., each client is itself a parameter server and synchronizes a global copy of parameters.
	
\subsection{Single Parameter Server} \label{sec:data:sps}

The Single Parameter Server strategy is the most naive parameter-server-based strategy for deep learning, with every GPU acting as a client and one worker GPU in charge of not only working but also synchronizing the parameters. In PyTorch and Tensorflow, this strategy is also the easiest to implement, which only requires tiny modifications of the original Python code, usually within less than 5 lines.
	
Before looking into the actual implementations, it's necessary to go through the abstract algorithm of SPS under the context of deep learning, which is shown in Algorithm \ref{algo:SPS}. The essence of this algorithm is to use one GPU as both the parameter server and worker, and all other GPUs as mere workers.\footnote{We can also interpret the other GPUs as clients that read and write data to the server.} In other words, only one GPU is used to store and synchronize the gradient needed for carrying on the training process. We represent the parameter server \texttt{GPU:0} as $G_0$ in the algorithm below. Note that this algorithm is batch-wise, which means that each time a data batch comes, this algorithm is repeated.

Before the algorithm starts, the model should be loaded onto the parameter server. The algorithm will broadcast the model with its parameters from the parameter server to each client, which replaces the original models on the clients. Then the system synchronizes until the model is fully established for all clients. After that, the parameter server scatters the data batch received and dispatches to each client for forwarding the loss. The system now waits for the calculation on each client to be finished. Up to this point, we have $n$ pieces of loss from $n$ GPUs, including the parameter server itself. At last, the parameter server $G_0$ gathers all the pieces of loss and concatenates them, and the concatenated loss object $J$ will go through the backward propagation process using the model on the parameter server and return the gradient results. Now the next data batch comes and this algorithm is executed once again.

\begin{algorithm}[htp!]
  \caption{SPS Data Parallel (for each batch)}
  \label{algo:SPS}
  \SetAlgoLined
  
  \KwData{model $M$, input tensor of size $[x, y, z]$, batch size $b$, GPUs $G_i$ where $i \in [0, n) \wedge \mathbb{N}$.}
  \KwResult{Result of gradient $W$ and loss $J$ after this batch.}
  \tcp{Now model M is on GPU:0}
  \For{$i \in [1, n)$}
  {
  	\textbf{broadcast}$(M, G_0, G_i)$; \\
  }
  \textbf{sync}$(G_0 \to G_n)$; \\
  \For{$i \in [0, n)$}
  {
  	$d_i$ = \textbf{scatter}$(b,x,y,z)$; \\
  	$j_i =$ \textbf{forward}$(d_i, G_i)$; \\
  }
  \textbf{sync} $(G_0 \to G_n)$; \\
  $J =$ \textbf{concat}$(j_0 \to j_{n-1}, G_0)$; \\
  $W =$ \textbf{backward}$(J, G_0)$; \\
 \Return{$J, W$}\;
\end{algorithm}

\begin{figure}[htp!]
	\begin{center}
  		\includegraphics[width=1.0\linewidth]{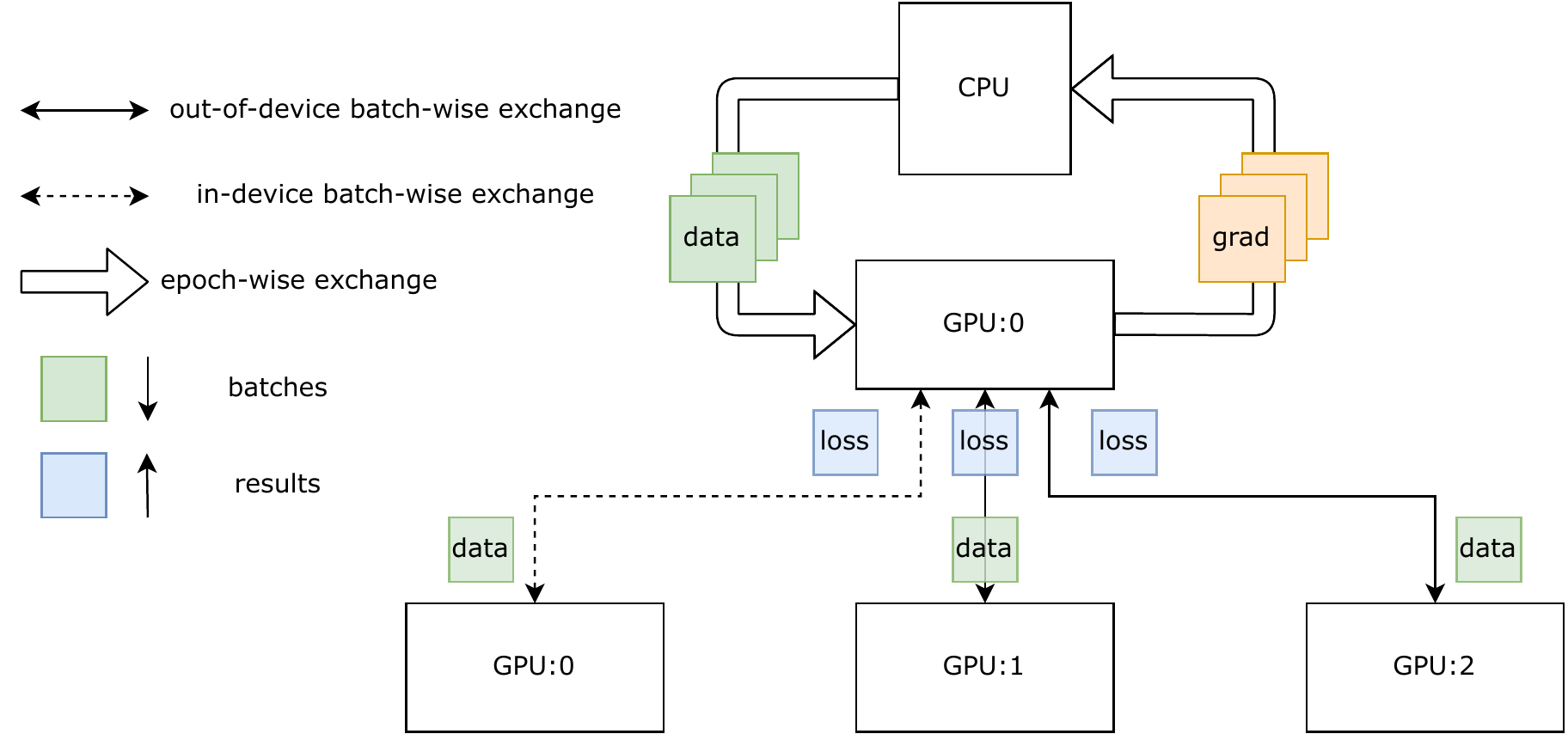}
	\end{center}
   	\caption{Conceptual architecture of Single Parameter Server parallization approach.}
	\label{fig:SPS}
\end{figure}

The conceptual illustration of the algorithm from a higher level is shown in Figure \ref{fig:SPS}. Each time the CPU sends an epoch of data to the parameter server \texttt{GPU:0}, the parameter server splits the epoch of data into batches, scatters each batch into pieces, and delivers each batch into a worker GPU for calculation. After synchronization of all GPUs, it aggregates the results and calculates the updated weight and loss. The system then repeats this process. After all the batches are processed, the result is delivered back to the CPU and begins calculating the next epoch.

The implementation of SPS is simple, consisting of selecting the parameter server, transferring the model to the parameter server, and wrapping the model up. The pseudo-code is shown below. 

\begin{lstlisting}[language={Python},numbers=left,numberstyle=\tiny,%frame=shadowbox,
   rulesepcolor=\color{red!20!green!20!blue!20},
   keywordstyle=\color{blue!70!black},
   commentstyle=\color{blue!90!},
   basicstyle=\ttfamily]
batch_size *= num_gpus
model = construct_training_model()
device = 'cuda:0'
model.to(device)
model = DataParallel(model, device_ids= 
    [list(range(num_gpus))])
\end{lstlisting}

Note that the new \texttt{batch\_size} in the code should be multiplied by the number of GPUs used, as the parameter server will split the batch into equivalent pieces for each GPU worker. Also, note that SPS does not support multi-node training, as data transfer among nodes is too slow if there is only one parameter server globally.\footnote{In the Distributed Parameter Server strategy, it's feasible to synchronize among nodes because only the gradients are being transferred, which is much smaller than the outputs. Also, because each GPU represents a parameter server, the communication is much faster than SPS.}

There is one variant implementation of SPS, and achievable in PyTorch. After profiling the performance, it has been found that the backward propagation process on the parameter server caused much overhead, because all the losses were concatenated to be processed on the parameter server. The solution is to leave this work to each client, and the parameter server should gather all the gradients (instead of losses) after each client finishes the gradient calculation. 

Although this implementation helps to improve the overall performance, this approach does not balance the load on each GPU either because each input batch is still scattered by the parameter server, which wastes time - the inequality of the GPUs always causes an unfair workload apportionment. To address this issue, the DPS strategy was proposed.

\subsection{Distributed Parameter Server} \label{sec:data:dps}

In the DPS strategy, each GPU holds the whole bunch of input batches, instead of waiting for the parameter server to scatter them. When a GPU finishes calculating the loss and gradient, it broadcasts them to all other GPU workers, as a result of all GPUs being parameter servers to this GPU.\footnote{There are many variants of the DPS strategies, mostly different in their number of parameter servers. In the context of this paper, we take the DPS strategy as a fully equally distributed parameter server system, where each GPU is both the parameter server and the client. Thus, each GPU worker stores a copy of parameters locally, and the system is fully decentralized.} Also, as the data sizes of loss and gradient are fairly small, the transmission overhead is greatly reduced. The algorithm is shown in Algorithm \ref{algo:DPS}.

In DPS, there are multiple CPU processes, each in charge of one GPU worker. Each process should be notified of how many total processes there are, which is called ``world size'', in order to make correct decisions. Each of them should also be notified of their order in the system, which is called ``rank'', to determine the synchronization order. If there are multiple nodes, the ``local rank'' is introduced to describe the rank of the process (or GPU) in the node it belongs to. As the system is now fully decentralized, there is no authorized arbitration for the scattering algorithm anymore. Thus, a protocol $P$ is required for all the processes to follow, which is predefined by PyTorch in its implementation.

Before the calculation for each batch begins, each GPU has one model that's already synchronized with all other GPUs. When the algorithm begins, all processes are delivered the same data batch, so each process firstly scatters the data batch using a pre-defined protocol, and gets its scattered data, which should be different from all others. Then each process stores the scattered data into the assigned GPU. When the GPU gets the data, it launches the training task immediately, calculating both the loss $j_i$ and gradient $w_i$. After this, the GPU transfers the intermediate results to its corresponding CPU immediately. Then the system waits for all the GPUs to finish the calculations, to achieve synchronization. When synchronized, one master process $C_m$ will gather all the loss and gradient pieces to calculate the final loss and gradient. 

The approach that the processes communicate with each other to calculate the final results is called all-reduce. This approach requires one master GPU to gather the intermediate result and calculate the final result. The final results will be broadcasted to all the other GPUs to update their model, so that each model remains synchronized \cite{patarasuk2009bandwidth}.
	
\begin{algorithm}[htp!]
  \caption{DPS Data Parallel (for each batch)}
  \label{algo:DPS}
  \SetAlgoLined
  
  \KwData{model $M$, input tensor of size $[x, y, z]$, batch size $b$, GPUs $G_i$ and CPUs $C_i$ where $i \in [0, n) \wedge \mathbb{N}$, scattering protocol $P$.}
  \KwResult{Result of gradient $W$ and loss $J$ after this batch.}
  \tcp{The model M is stored on each GPU}
  \For{$i \in [0, n)$}
  {
  	$d_i =$ \textbf{scatter}$(b,x,y,z,P, C_i)$; \\
  	\textbf{store}$(d_i, G_i)$; \\
  	$j_i =$ \textbf{forward}$(d_i, G_i)$; \\
  	$w_i =$ \textbf{backward}$(j_i, G_i)$; \\
  	\textbf{store}$(j_i, w_i, C_i)$; \\
  	\textbf{sync} $(G_0 \to G_n)$; \\
  	$W,J = $ \textbf{allreduce}$(w_i, j_i, C_m)$; \\
  }
  \Return{$J, W$}\;
\end{algorithm}

\begin{figure}[htp!]
	\begin{center}
  		\includegraphics[width=1.0\linewidth]{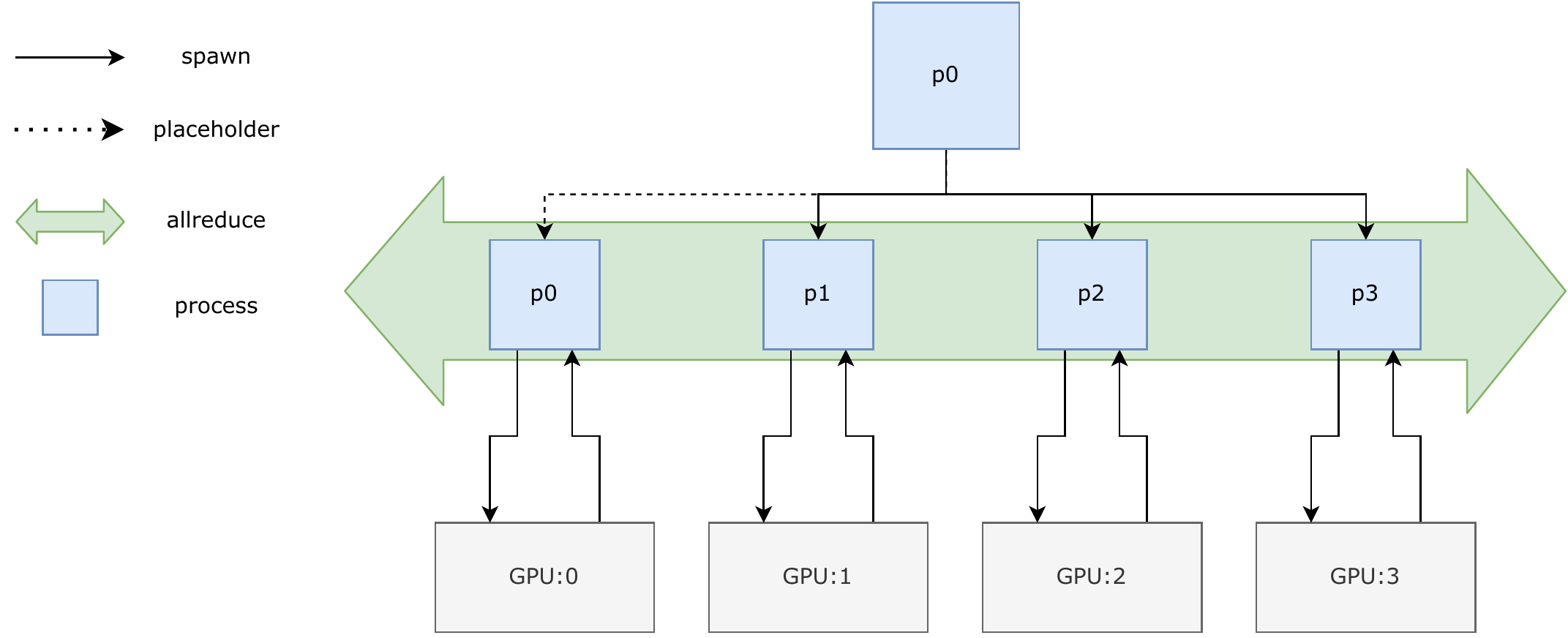}
	\end{center}
   	\caption{Conceptual architecture of Distributed Parameter Server parallization approach.}
	\label{fig:DPS}
\end{figure}

The conceptual architecture of the DPS strategy is shown in Figure \ref{fig:DPS}. After the CPU launches a master process $p_0$, the master process automatically spawns enough processes, so that each process will be in charge of one GPU. Each process gets the same copy of data, but they scatter the data differently, according to the pre-defined protocol. Then each CPU is in charge of one GPU to process the learning process, and use the allreduce functionality to synchronize the parameters.

Different from the simplicity when applying SPS, implementing DPS is more complex. For launching multiple processes, the \texttt{MultiProcessing} library should be imported. For the initialization part, if the process is going to run on multiple nodes, we should launch the training script on all of the nodes. Moreover, we need to tell all the processes which one is the top process that launches other processes. The detailed Python script is shown below.
	
\begin{lstlisting}[language={Python},numbers=left,numberstyle=\tiny,%frame=shadowbox,
   rulesepcolor=\color{red!20!green!20!blue!20},
   keywordstyle=\color{blue!70!black},
   commentstyle=\color{blue!90!},
   basicstyle=\ttfamily]
arg.world_size = arg.gpus * arg.nodes
os.environ['MASTER_ADDR'] = '..'
os.environ['MASTER_PORT'] = '..'
mp.spawn(train, nprocs=args.gpus, 
    args=(args,))
\end{lstlisting}
	
We also need to modify the training function to fit the code into the DPS strategy. In the training function \texttt{train(gpu, args)}, before wrapping the model, we need to calculate the rank number of each process and initialize the process group. Then we need to notify all processes in the process group where to find the global settings like \texttt{world\_size}, using \texttt{init\_method}. What's more, we need to use the distributed scatterer to split each batch of data into pieces with a protocol, so that their scattered data pieces don't overlap. The pseudo-code is shown below.

\begin{lstlisting}[language={Python},numbers=left,numberstyle=\tiny,%frame=shadowbox,
   rulesepcolor=\color{red!20!green!20!blue!20},
   keywordstyle=\color{blue!70!black},
   commentstyle=\color{blue!90!},
   basicstyle=\ttfamily]
rank = args.nr * args.gpus + gpu	                          
dist.init_process_group()
model = construct_training_model()
model = DistrDataParal(model, ..)
train_dataset = construct_dataset()
train_sampler = DistrSampler()
train_loader = DL()
\end{lstlisting}

In order to make the new version work, the user should launch the code on each working node. For example, if there are 4 nodes available, the user should use the following command on each node: 

\begin{lstlisting}[language={Python},numbers=left,numberstyle=\tiny,%frame=shadowbox,
   rulesepcolor=\color{red!20!green!20!blue!20},
   keywordstyle=\color{blue!70!black},
   commentstyle=\color{blue!90!},
   basicstyle=\ttfamily]
python main.py -n 4 -g 8 -nr i
\end{lstlisting}

In this command, $n$ is the total number of the computation nodes, $g$ is the total number of GPUs, and $i$ is the node rank, where $i\in \{0, 1, 2, 3\}$. Note that batch size after applying the DPS strategy is the world size multiplied by \texttt{batch\_size} shown in the code above, as each GPU is designated for the same amount of \texttt{batch\_size}. This is different from the \texttt{batch\_size} in the SPS strategy, where it means the total batch size for the whole system.

If only one node is required for data-parallelization, the launch script \texttt{torch.launch} provided by Python can be a shortcut.\footnote{Note that this approach is only valid when we are running on one node. On multiple nodes, it's still required to use the approach mentioned above.} To lunch the simplified script, it's required to add some more arguments when executing. For example, when the program uses 4 GPUs on the node, the following command should be executed.

\begin{lstlisting}[language={Python},numbers=left,numberstyle=\tiny,%frame=shadowbox,
   rulesepcolor=\color{red!20!green!20!blue!20},
   keywordstyle=\color{blue!70!black},
   commentstyle=\color{blue!90!},
   basicstyle=\ttfamily]
python -m torch.distributed.launch 
  --nproc_per_node=4 main.py
\end{lstlisting}

The \texttt{train\_sampler}, \texttt{train\_loader}, and model wrapping parts should be the same as the multiple nodes multiple GPUs approaches. There are also some other minor fix when using the launch script, which has been gathered on the GitHub repo.\footnote{The repository locates at \url{https://github.com/BiEchi/DistributedTrainingGPT2/tree/main/GPT2-Chinese-Parallel-Distributed}.}

There are also multiple implementations of this strategy. For example, the allreduce algorithm still needs a master process to calculate the final loss and gradient, which cannot be applied to large-scale tasks, because the calculation linearly expands when the number of workers increases. Thus, some DPS strategies replace the allreduce algorithm with the ring allreduce algorithm, which will be illustrated in Section \ref{sec:data:hvd}. Other variants also try to synchronize the losses and gradients directly through the GPUs instead of going through CPUs, thus saving time due to storing the data back to the CPUs.

Although DPS is a very powerful approach to use on one node, for multiple nodes, it's far from convenient because it requires each node to execute a script. What's more, PyTorch implements the allreduce algorithm version of the DPS strategy, which is not scalable, as mentioned in the last paragraph. Thus, a bit later than the invention of DPS, the Horovod framework was invented to utilize the MPI backend for inter-node communication, which saves time to modify, and minimizes the data transmission overhead.
	
\subsection{The Horovod Framework For DPS} \label{sec:data:hvd}

Horovod is a very popular distributed deep learning framework supporting Tensorflow, Keras, PyTorch, and lots of other deep learning frameworks \cite{sergeev2018horovod}. Different from the original DPS algorithm, Horovod utilizes the ring allreduce algorithm for communication among workers.

\begin{figure}[htp!]
	\begin{center}
  		\includegraphics[width=0.9\linewidth]{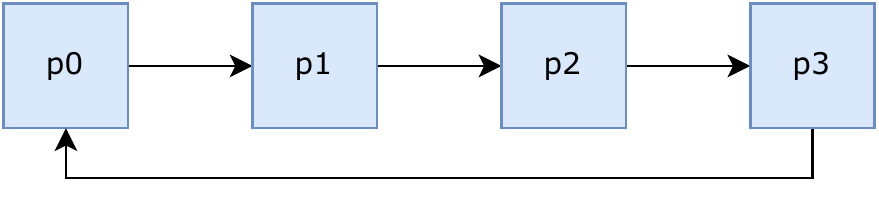}
	\end{center}
   	\caption{The conceptual schema of the Ring Allreduce algorithm.}
	\label{fig:ring-allreduce}
\end{figure}

The Ring Allreduce algorithm groups all GPUs into a ring topology, with each GPU abstracted as a node \cite{hillis1986data}, which is illustrated in Figure \ref{fig:ring-allreduce}. If there are $n$ processes in   total, then when synchronizing the parameters, each process will divide its gradient into $n$ blocks, then sends this piece of gradient to the next GPU in the ring for synthesizing. The details of the algorithm are illustrated in the paper proposing Horovod \cite{hillis1986data}, where this algorithm is also proved to be a bandwidth-optimized algorithm.

Different from the DPS approach in PyTorch, Horovod uses the external backend \texttt{openmpi} for communication among the nodes in the network, so before using Horovod, it's required to install MPI on each computing node. After installing MPI, the node that launches Horovod will be able to communicate with all other nodes requested to run Horovod using the \texttt{openmpi} toolkit. The Python implementation of Horovod is basically the same as DPS except for minor fixes, which is shown below.

\begin{lstlisting}[language={Python},numbers=left,numberstyle=\tiny,%frame=shadowbox,
   rulesepcolor=\color{red!20!green!20!blue!20},
   keywordstyle=\color{blue!70!black},
   commentstyle=\color{blue!90!},
   basicstyle=\ttfamily]
hvd.init()
local_rank = hvd.local_rank()
torch.cuda.set_device(local_rank)
train_sampler = DistributedSampler()
train_loader = DataLoader()
optimizer = optim.SGD()
optimizer = hvd.DistributedOptimizer()
hvd.broadcast_parameters()
hvd.broadcast_optimizer_state()
device = torch.device()
\end{lstlisting}

To run Horovod on one node, simply use the command \texttt{horovodrun -np 4 -H localhost:4 python train.py} if there are 4 GPUs on the node. To run Horovod on multiple nodes, it's required to explicitly list the node name in the launch command. For example, \texttt{horovodrun -np 8 -H nodename1:4,nodename2:4 python train.py} will launch a task with 8 processes, and \texttt{nodename\{n\}} is the name of the node in the cluster network, like \texttt{hal08}.

\begin{table*}
\parbox{0.475\linewidth}{
\centering
\setlength\tabcolsep{2.3pt}
\small
\begin{tabular}{c|c|cccccc}
\hline
\# Parallel Type & \# PrmSrv & \# MaxBatch & Process & Node & GPU & Tor & TF \\
\hline
Baseline & 0 & 4 & 1 & 1 & 1 & Y & Y\\
\hline
SPS & 1 & 16 & 1 & 1 & 4 & Y & Y\\
DPS & 4 & 2*4 & 4 & 1 & 4 & Y & Y\\
HVD & 4 & 4*4 & 4 & 1 & 4 & Y & Y\\
HVD-Apex & 4 & 4*4 & 4 & 1 & 4 & Y & Y\\
HVD-Multi & 8 & 4*8 & 8 & 2 & 8 & Y & Y\\
\hline
\end{tabular}
\caption{Behavior of each data parallel architecture with respect to their PrmSrv (number of parameter server), MaxBatch(maximum batch size) supported on \textit{HAL} V100 GPU groups, Process number, and whether they're supported by PyTorch, Tensorflow. Y means the strategy is supported by the official documentation. \label{tab:capability}}
}
\hfill
\parbox{0.475\linewidth}{
\centering
\setlength\tabcolsep{2.3pt}
\small
\begin{tabular}{c|c|ccccc}
\hline
Parallel Type & PrmSrv & GPU0 & GPU1 & GPU2 & GPU3 & Total \\
\hline
Baseline & 0 & 85\% & 0 & 0 & 0 & 0.94 \\
SPS & 1 & 17\% & 13\% & 11\% & 14\% & 0.55 \\
\hline
DPS & 4 & 94\% & 94\% & 89\% & 94\% & \underline{3.71} \\
DPS-Apex & 4 & 87\% & 88\% & 94\% & 92\% & 3.61 \\
DPS-ClrMem & 4 & 52\% & 38\% & 44\% & 29\% & 1.63 \\
\hline
Horovod & 4 & 92\% & 86\% & 85\% & 81\% & 3.44 \\
Horovod-Apex & 4 & 84\% & 90\% & 91\% & 90\% & 3.55 \\
\hline
Horovod-Multi & 8 & 182\% & 175\% & 162\% & 163\% & 6.82 \\
Horovod-MA & 8 & 188\% & 181\% & 174\% & 171\% & \textbf{7.14} \\
\hline
\end{tabular}
\caption{GPU utilization of different distributed approaches. \textbf{Bold}: global best performance. \underline{Underlined}: best performance using one node. \label{tab:gpu_usage}}
}
\end{table*}

\subsection{Apex Half Precision} \label{sec:explain-apex}

Strictly speaking, The Apex half-precision training is not a distributed deep learning strategy per se, but a crucial tool to accelerate deep learning. The half-precision training can work without distributed deep learning, but can be integrated into the DPS and Horovod frameworks after minor changes to code, which accelerates the learning rate to an even higher extent. Thus, this research also takes Apex half-decision training into consideration.

The half-precision (\texttt{fp16}) training is often mixed with single precision (\texttt{fp32}) training. With the introduction of the half-precision floating numbers, GPUs take less storage\footnote{For a more detailed effect of half-precision numbers have on memory usage, please turn to Appendix \ref{sec:apex}.} and time to process the data so the batch size can be incremented. Half precision is also capable of utilizing NVIDIA Tensor Cores to further accelerate computing efficiency. 

However, only GPUs supporting mixed precision calculations can make use of this strategy. To integrate \texttt{fp16} into any framework, the first change is to use the backward function provided by Apex \texttt{amp} library. For either SPS or DPS based on the PyTorch framework, the modifications are shown below.

\begin{lstlisting}[language={Python},numbers=left,numberstyle=\tiny,%frame=shadowbox,
   rulesepcolor=\color{red!20!green!20!blue!20},
   keywordstyle=\color{blue!70!black},
   commentstyle=\color{blue!90!},
   basicstyle=\ttfamily]
model, optimizer = amp.initialize()
model = DDP(model)
with amp.scale_loss() as scaled_loss: 
    scaled_loss.backward()
\end{lstlisting}
	
Horovod does not officially support \texttt{amp}, but a solution provided by the community is available, with the modification shown below.

\begin{lstlisting}[language={Python},numbers=left,numberstyle=\tiny,%frame=shadowbox,
   rulesepcolor=\color{red!20!green!20!blue!20},
   keywordstyle=\color{blue!70!black},
   commentstyle=\color{blue!90!},
   basicstyle=\ttfamily]
with amp.scale_loss() as scaled_loss:
    scaled_loss.backward()
    if FLAGS_horovod:
        optimizer.synchronize()
        with skip_synchronize():
            step()
\end{lstlisting}

Another note is that as \texttt{fp16} does not provide good performance improvement under small batches (as shown in Section \ref{sec:experiment}) due to already quick training time per step brought by the small batch size, it's not recommended to turn on \texttt{fp16} when the batch is small. Another concern with using Apex under small batches is that, due to the mechanism of half-precision calculations, the training may face a gradient overflow issue, which decreases the training performance.

\section{Experiments} \label{sec:experiment}
In this part, we show the experimental results of all the strategies mentioned in Section \ref{sec:data}. The developed code is available on GitHub Repository \url{https://github.com/BiEchi/DistributedTrainingGPT2}. 

In our experiments, we utilized the GPT2-Chinese model taken from GitHub Repo \url{https://github.com/Morizeyao/GPT2-Chinese} as the starting point. The hyperparameters are listed in Table \ref{tab:hypers},\footnote{The hyperparameters are also available directly in code at \url{https://github.com/BiEchi/DistributedTrainingGPT2/blob/main/GPT2-Chinese-Parallel/config/model_config.json}} and the dataset is simply a list of sentences extracted from a novel.

\begin{center}
\begin{table}[htp!]
\centering
\small
\setlength\tabcolsep{2.3pt}
\begin{tabular}{c|c|c}
\hline
Hyper-Parameter & Marker & Size\\
\hline
\# embedded layers & \texttt{n\_embd} & $768$ \\
\# heads per attention layer & \texttt{n\_head} & $12$ \\
\# layers & \texttt{n\_layer} & $12$ \\
vocabulary size & \texttt{vocab\_size} & $26679$ \\
\hline
\end{tabular}
\caption{Selected hyper-parameters and their values.}
\label{tab:hypers}
\end{table}
\end{center}

\subsection{Qualitative Results}

In this part, we use Table \ref{tab:capability} to show the arguments of each data-parallel strategy after experiments and the framework support (PyTorch and Tensorflow) for the strategies.

From the table, we see that all the data parallel types are officially supported by both PyTorch and Tensorflow. The baseline model has no parameter servers, the SPS strategy utilizes only one parameter server, and the DPS and Horovod strategies support multiple parameter servers. If the strategy is parameter-server-based, the number of processes should be the same as the number of parameter servers, because each parameter server requires an individual process to launch. More insights into other columns in this table will be introduced after Section \ref{sec:graphical}.

\subsection{Quantitative Results} \label{sec:graphical}

In this part, we analyze the GPU utilization and time consumption for each data-parallel strategy.

The GPU utilization for each strategy is shown in Table \ref{tab:gpu_usage}. Note that Horovod with multiple nodes has much higher GPU usage because we sum up the usage in two nodes. From the table, we conclude that under the limitation of 1 node, DPS has the overall best performance, and Horovod is only a little below it, while SPS performs much worse than them, with the GPU utility even worse than the baseline. 
	
We then run the data parallelization on GPT-2 with a parameter amount of 10274200 (10M, similar to GPT2-mini) and 106310400 (100M, similar to GPT2-small), respectively. The time taken by different strategies is shown in Table \ref{tab:efficiency}. Note that we don't run GPT2-base for any strategies with only one node because it takes more than 24 hours to finish.

\begin{center}
\begin{table}[htp!]
\centering
\small
\setlength\tabcolsep{2.3pt}
\begin{tabular}{c|c|cc}
\hline
Parallel Type & PrmSrv & 10M (mini) & 100M (small)\\
\hline
Baseline & 0 & $174m$ & $>24h$ \\
\hline
SPS & 1 & $186m$ & $>24h$ \\
DPS-Est & 4 & $26m$ & $228m$ \\
DPS-Apex & 4 & $14m$ & $133m$ \\
Horovod & 4 & $42m$ & $451m$ \\
Horovod-Apex & 4 & $22m$ & $198m$  \\
Horovod-Multi & 8 & $28m$ & $292m$ \\
Horovod-MA & 8 & $18m$ & $162m$ \\
\hline
\end{tabular}
\caption{Efficiency of different parallel patterns. Horovod-Multi means multi-node Horovod strategy, and Horovod-MA means Multi-node Horovod with Apex mix precision training.}
\label{tab:efficiency}
\end{table}
\end{center}

We use the hooks provided by PyTorch to record the loss curve of each strategy with respect to the time or steps during training. 

    \begin{figure*}
        \centering
        \begin{subfigure}[b]{0.475\textwidth}
            \centering
            \includegraphics[width=\textwidth]{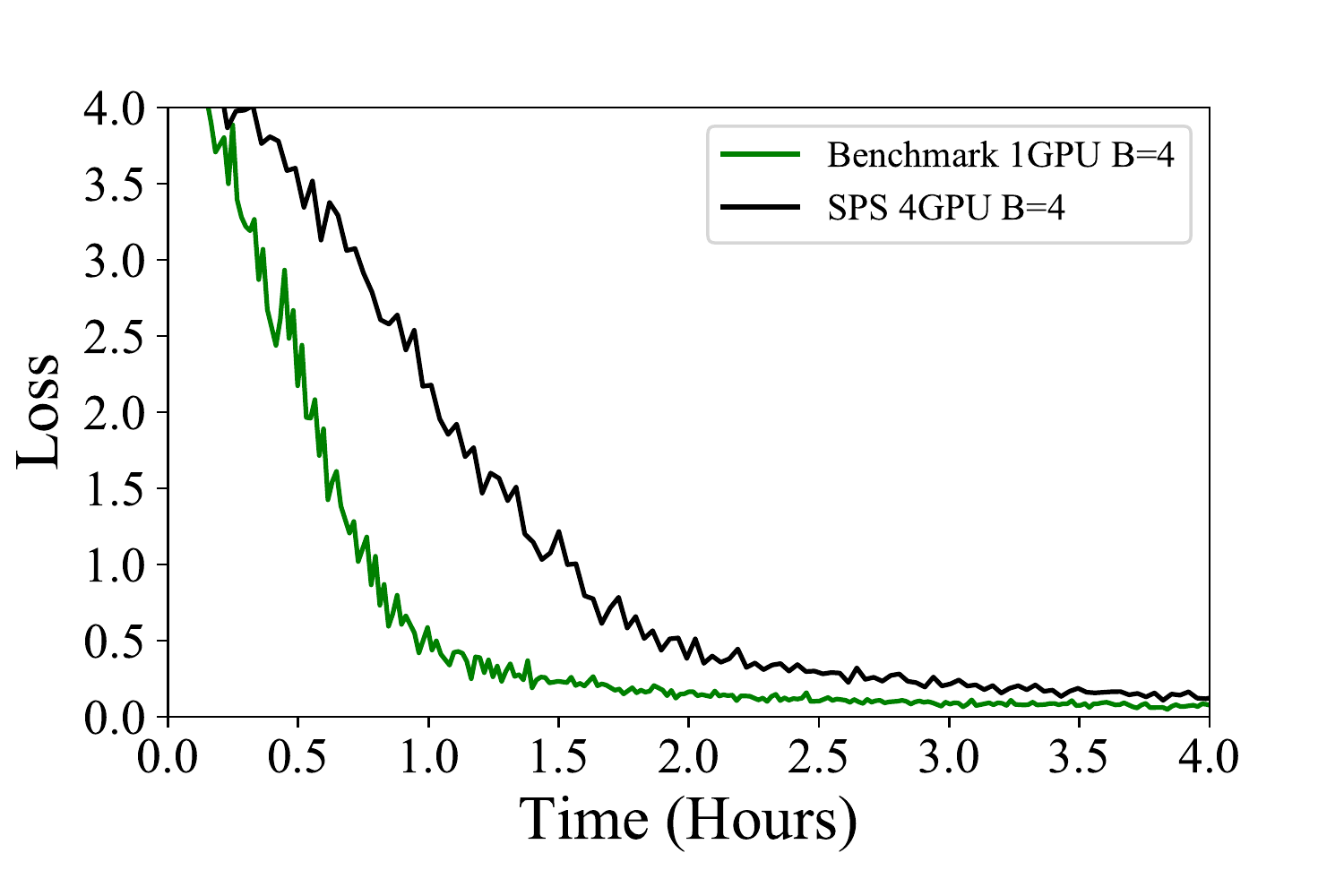}
            \caption[]%
            {{\small Loss-time curve for benchmark and SPS with batch size of 4. Under the same condition the benchmark is about 3 times faster than SPS solution.}}    
            \label{fig:bm-sps}
        \end{subfigure}
        \hfill
        \begin{subfigure}[b]{0.475\textwidth}  
            \centering 
            \includegraphics[width=\textwidth]{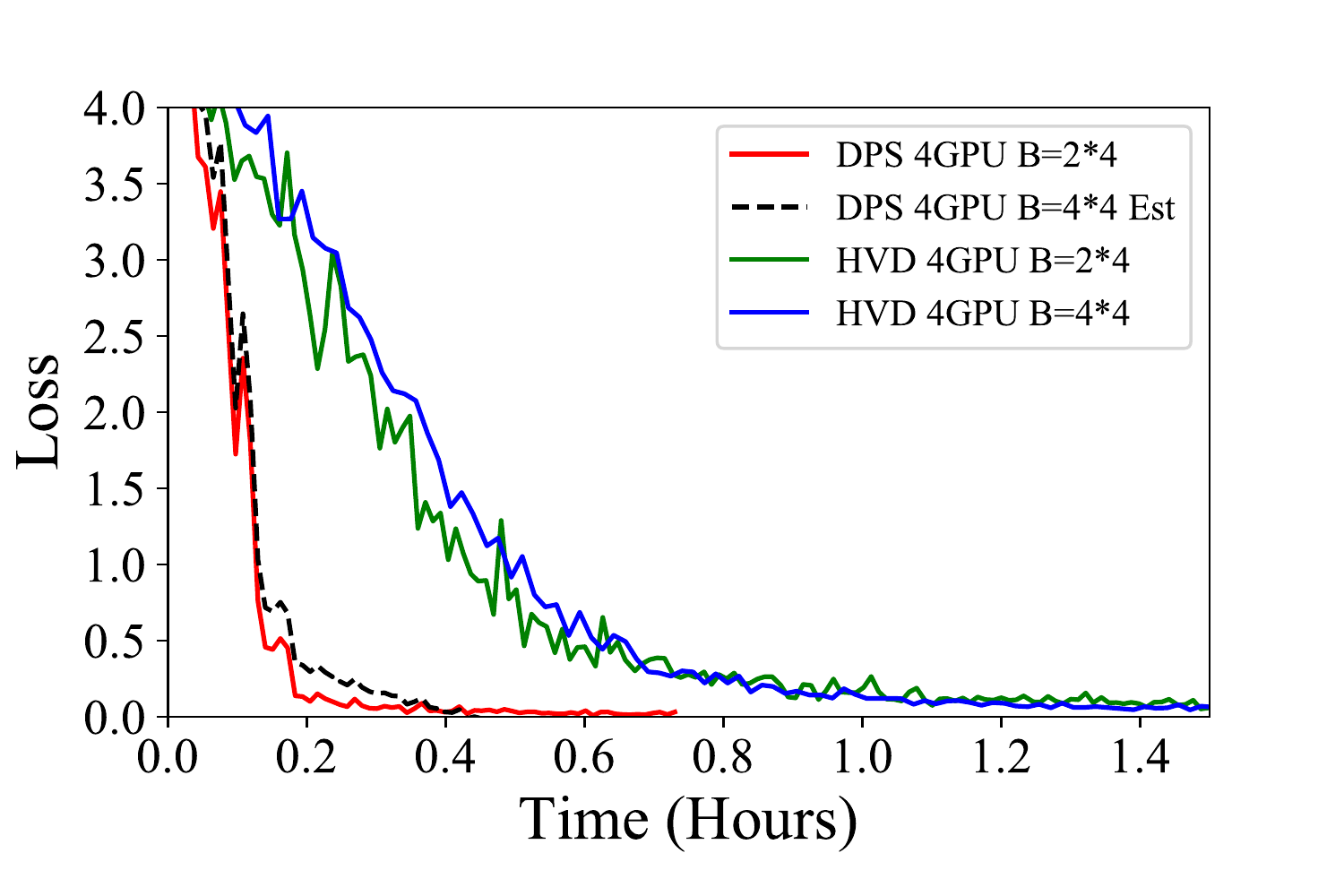}
            \caption[]%
            {{\small Loss-time curve for DPS and Horovod with batch sizes of 8 and 16. The dashed black line indicates an estimated loss curve using the same tendency as Horovod.}}    
            \label{fig:dps-hvd-2-4}
        \end{subfigure}
        \caption[]
        {\small Experimental results of SPS compared to the benchmark (a), and DPS compared to SPS (b).}
        \label{fig:bm-sps-dps-hvd}
    \end{figure*}

The loss curve of the SPS strategy (4 GPUs, total batch size 4) and benchmark (1 GPU, batch size 4) with respect to training time is shown in Figure \ref{fig:bm-sps}. Recall that if the batch size of the SPS solution is 4, each GPU will be assigned only 1 batch. According to the result, under the same batch size, the performance of SPS with 4 GPUs is even worse than the benchmark with only 1 GPU.

Referring back to Section \ref{sec:data:sps}, the probable cause is that the back propagation on the parameter server takes too much time. As shown in Algorithm \ref{algo:SPS}, the workers can not go on with the next batch of input until the parameter server finishes the back propagation process. This can be further proved when we refer to Table \ref{tab:gpu_usage}: the GPU utilization of SPS is extremely low, and we see some phases when only the parameter server is in use.

\begin{figure*}
        \centering
        \begin{subfigure}[b]{0.475\textwidth}
            \centering
            \includegraphics[width=\textwidth]{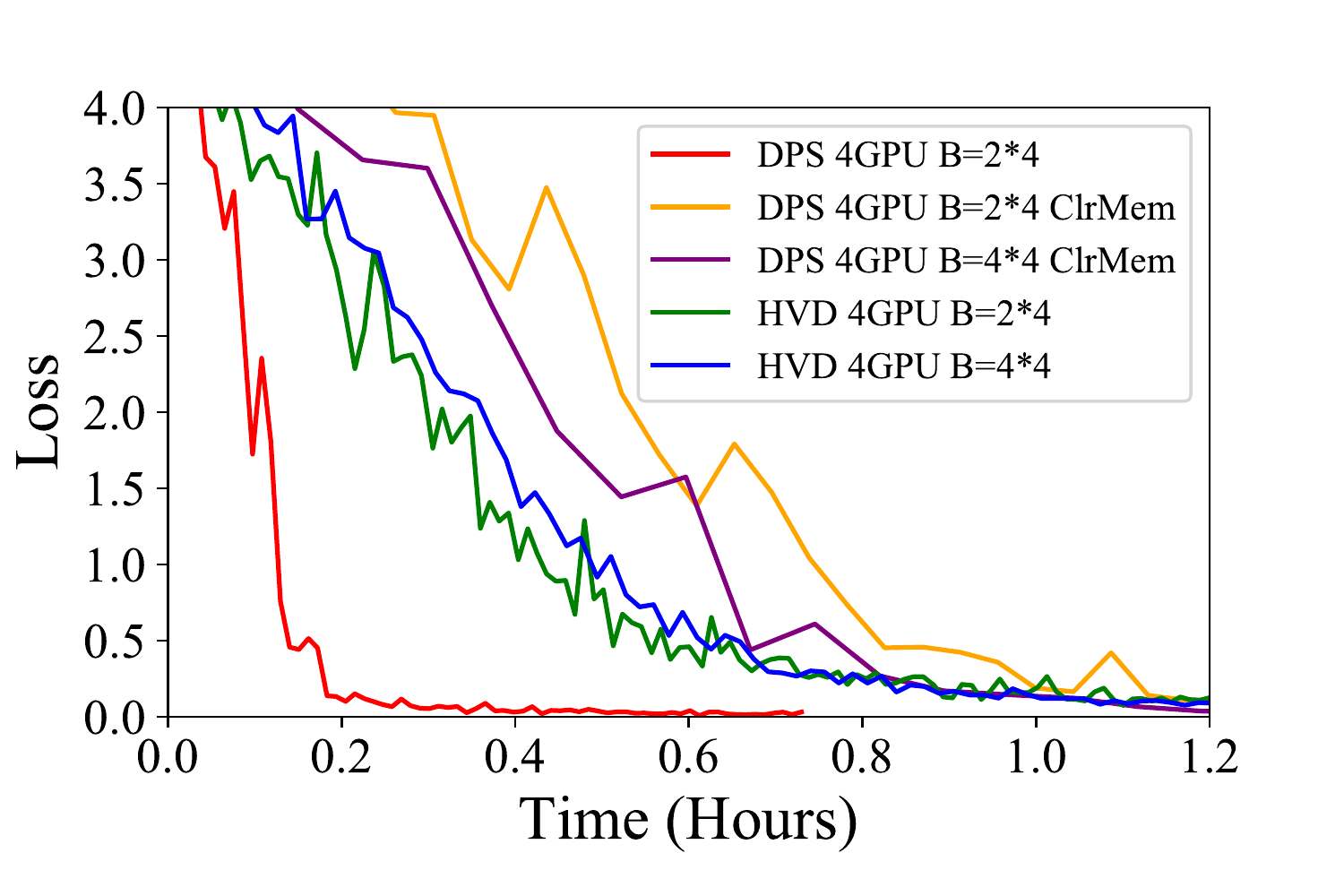}
            \caption[]%
            {{\small Loss curve for DPS (after adding the memory cleaning code) and Horovod, both with respect to a batch size of 8 and 16.}}    
            \label{fig:dps-clrmem}
        \end{subfigure}
        \hfill
        \begin{subfigure}[b]{0.475\textwidth}  
            \centering 
            \includegraphics[width=\textwidth]{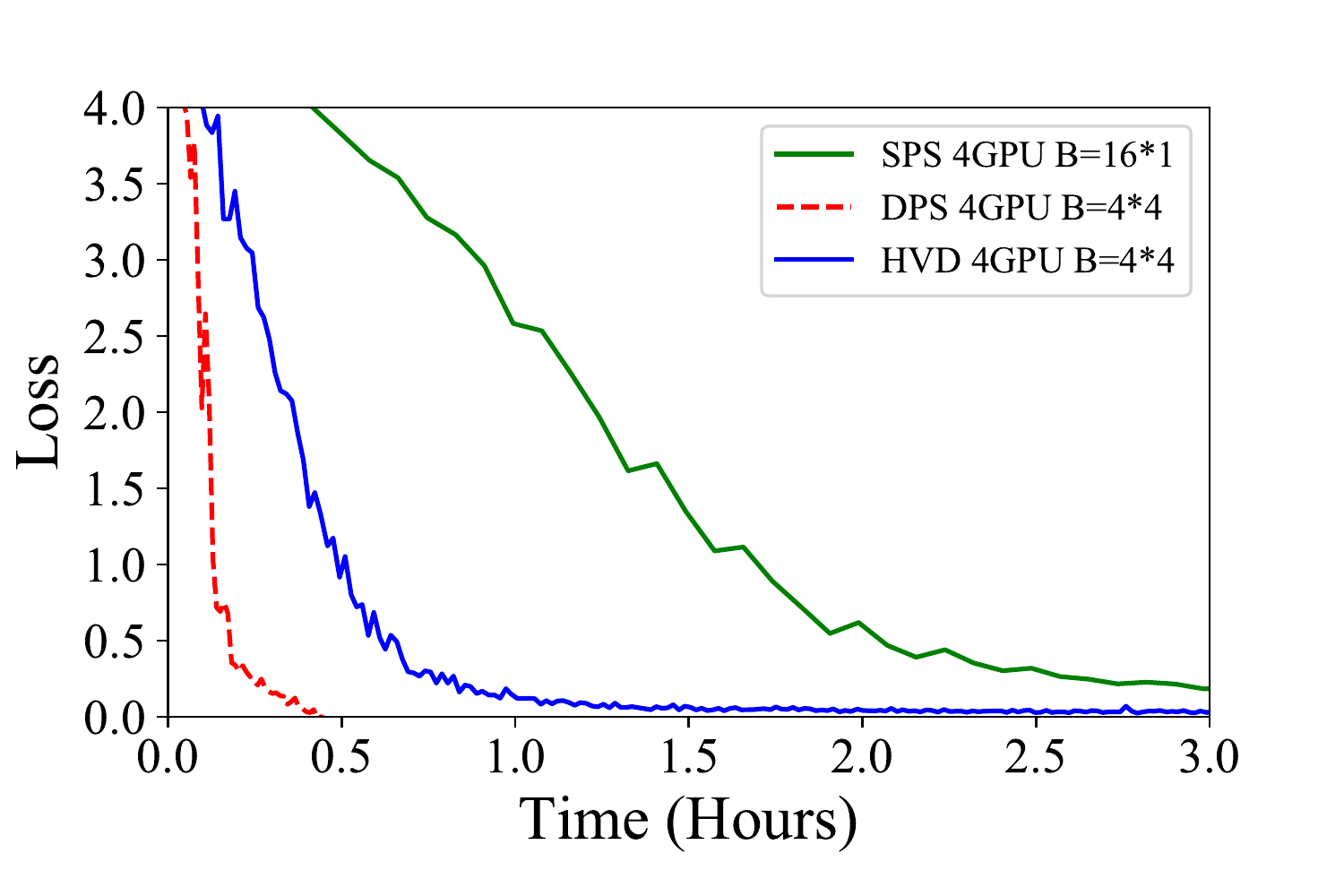}
            \caption[]%
            {{\small Loss curve for SPS, DPS and Horovod under the same batch size (16).}}    
            \label{fig:sps-dps-hvd}
        \end{subfigure}
        \vskip\baselineskip
        \begin{subfigure}[b]{0.475\textwidth}   
            \centering 
            \includegraphics[width=\textwidth]{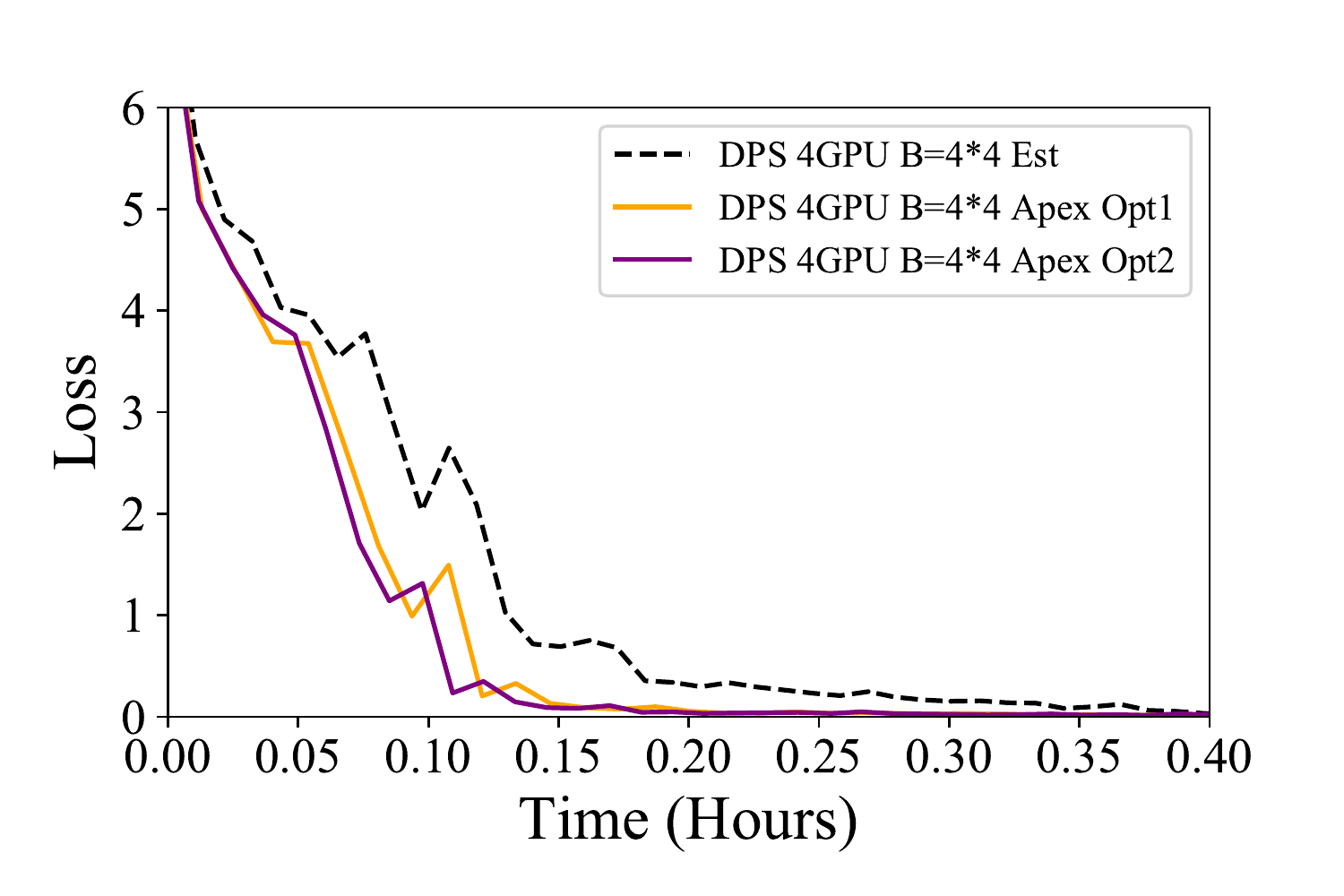}
            \caption[]%
            {{\small Loss curve for DPS before and after Optimization 1 \& 2 (\texttt{O1} \& \texttt{O2}) using Apex \texttt{fp16} mixed training.}}    
            \label{fig:dps-dpsapex}
        \end{subfigure}
        \hfill
        \begin{subfigure}[b]{0.475\textwidth}   
            \centering 
            \includegraphics[width=\textwidth]{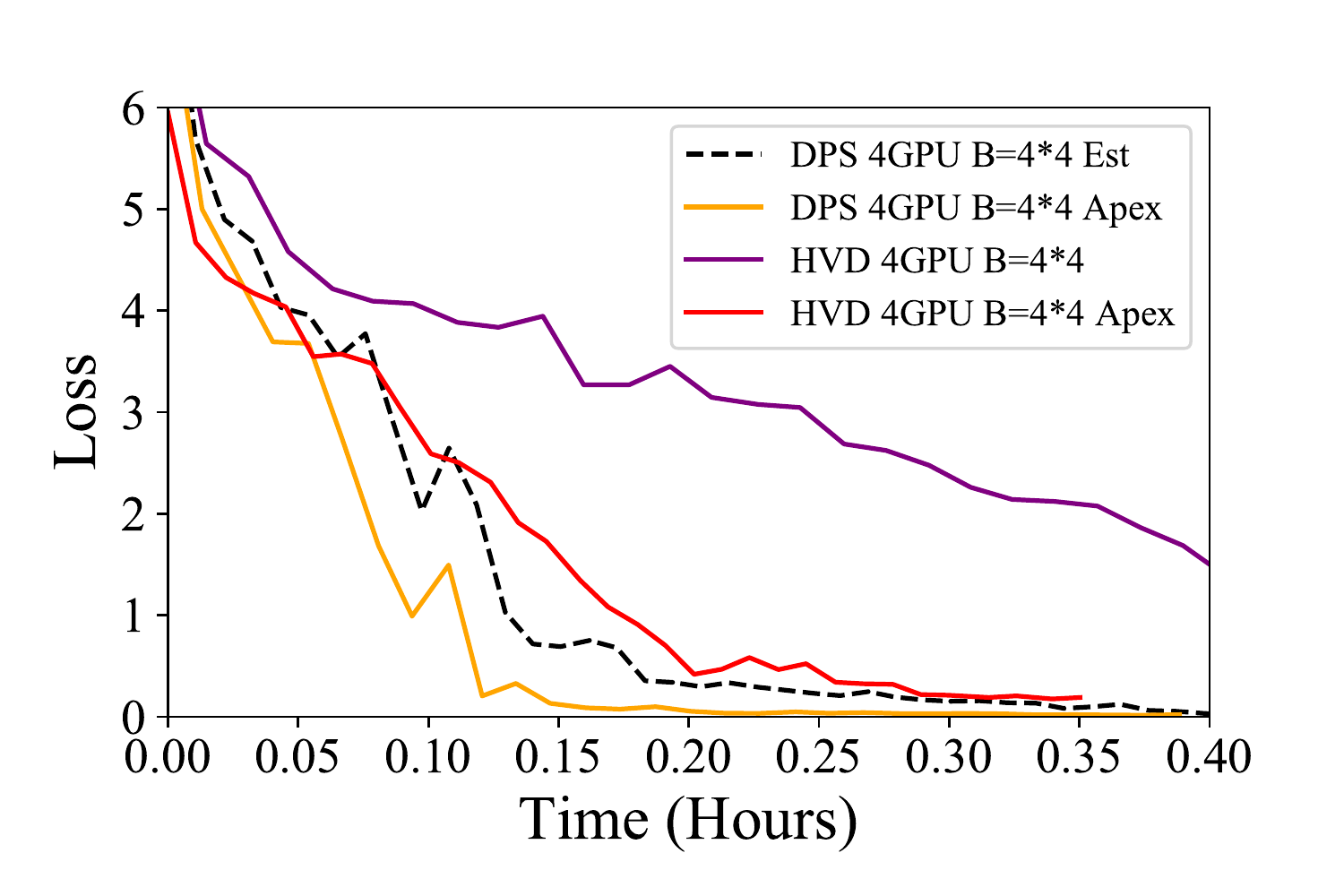}
            \caption[]%
            {{\small Loss curve for DPS and Horovod before and after applying Apex \texttt{fp16-fp32} mixed training.}}    
            \label{fig:hvd-hvdapex}
        \end{subfigure}
        \caption[]
        {\small Experimental results of DPS with/without cleaning memory compared to Horovod (a), horizontal comparison of SPS, DPS and Horovod (b), performance difference in optimization of DPS (c), and DPS and Horovod before/after applying Apex (d).}
        \label{fig:dps-hvd-4graph}
    \end{figure*}

We then move on to train with the same arguments using the DPS strategy and the Horovod strategy. The results of SPS, DPS, and Horovod under a total batch size of 16 are shown in Figure \ref{fig:sps-dps-hvd}. During our training process, we unexpectedly found that DPS runs out of memory under the batch size of 4*4 (4 GPUs, each with batch size 4), while Horovod works well. The reason is related to the difference between the implementation of DPS and Horovod, more on this in Appendix \ref{sec:data-parallel}. Therefore, we decreased the batch size to 2 for both DPS and Horovod to get the training process to work. The results are shown in Figure \ref{fig:dps-hvd-2-4}. It's discovered that it costs slightly more time for Horovod to train on a batch size of 4 than that of 2. Based on this, we estimated the loss function of DPS under a batch size of 4 to be a bit higher than that of 2, as illustrated in the figure. For a detailed analysis of the cause of CUDA OOM errors, please turn to Appendix \ref{sec:rules-gpu-mem} and \ref{sec:reduce-gpu-mem}.

This issue implies the scalability of Horovod. As mentioned in Section \ref{sec:data:hvd}, Horovod uses the ring-allreduce strategy for communication among the workers, which totally gets rid of the inequality of the GPU workers. It's also discovered that Horovod includes built-in functionalities like cleaning memory caches and balancing the workload on each GPU. 

According to the results, it takes approximately half more time for Horovod to finish the training task, probably due to the framework overhead of Horovod, including the memory cleaner mentioned. However, the power of Horovod should not be underestimated: according to Figure \ref{fig:dps-clrmem}, after we implemented the garbage cleaner code for DPS, the training efficiency with respect to time decreased by a factor of 3. Although the codes added are simply \texttt{del loss} and \texttt{empty\_cache()}, it causes a huge overhead for memory cleaning, which shows the value of Horovod. For the analysis of the effect of the garbage cleaning on the training performance, refer to Appendix \ref{sec:free-vars}.

Based on these observations, it can be pointed out that DPS takes the first place due to its extremely high performance on one node, while it does not deal with memory cleaning and workload balancing and may cause the code to run out of memory. However, both DPS and Horovod significantly outperform SPS.

    \begin{figure*}
        \centering
        \begin{subfigure}[b]{0.475\textwidth}
            \centering
            \includegraphics[width=\textwidth]{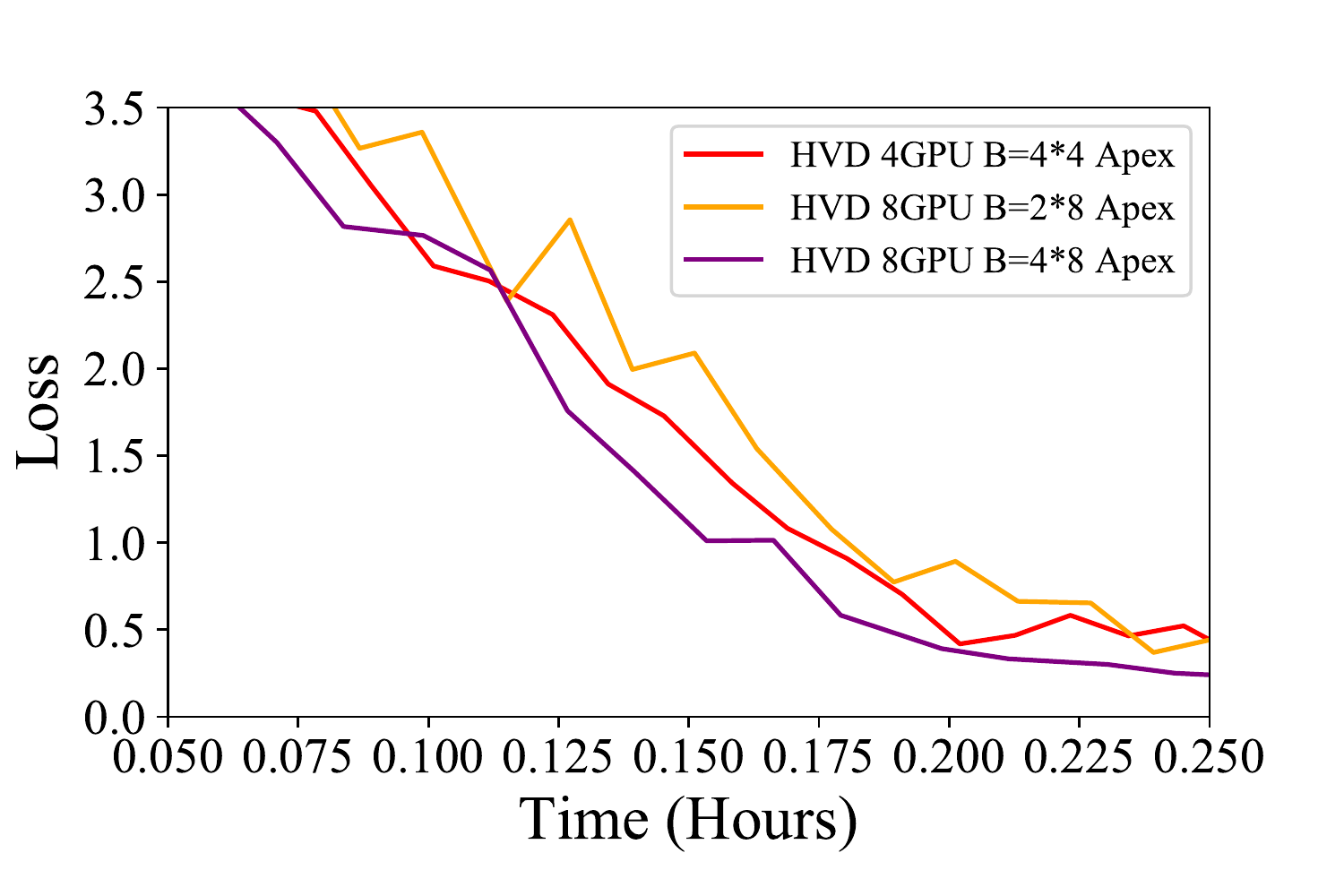}
            \caption[]%
            {{\small Loss curve for Horovod-Apex under 1 node (4 GPUs) and 2 nodes (8 GPUs) under batch sizes of 4*4, 2*8 and 4*8.}}    
            \label{fig:hvdapex-4-8}
        \end{subfigure}
        \hfill
        \begin{subfigure}[b]{0.475\textwidth}  
            \centering 
            \includegraphics[width=\textwidth]{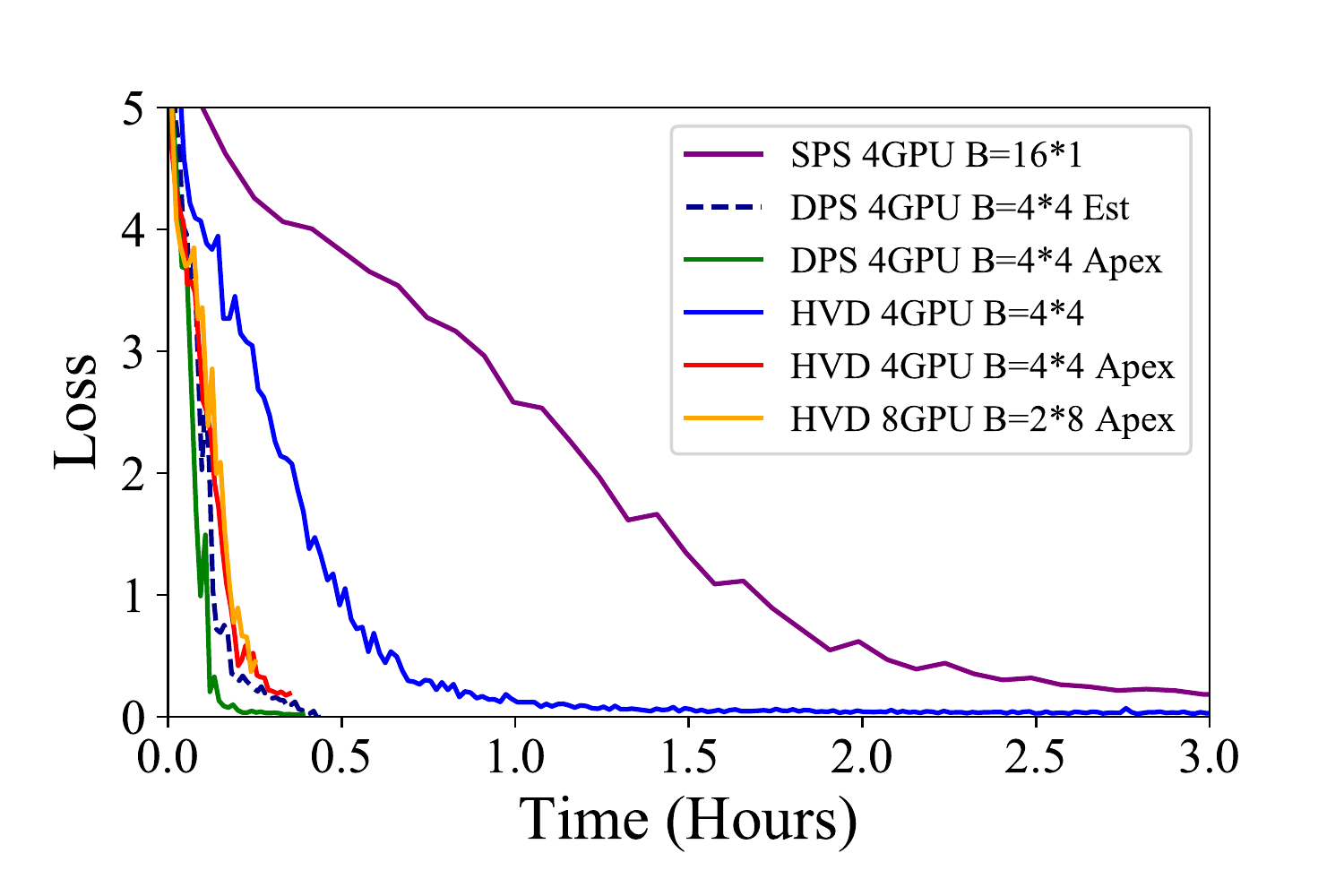}
            \caption[]%
            {{\small Loss curve for all basic training approaches, including SPS, DPS and Horovod.}}    
            \label{fig:all}
        \end{subfigure}
        \caption[]
        {\small The experimental results of single/multiuple node training using Horovod-Apex (a), and a comprehensive comparison of all the optimizations concerned (b).}
        \label{fig:hvdapex-all}
    \end{figure*}

We also tested the performance of DPS with Apex and Horovod with Apex. We did not take SPS into consideration because it's ten-fold slower. We first implemented Apex on DPS using 4 GPUs and a batch size of 4*4 to probe whether the overall mechanism works. The result is shown in Figure \ref{fig:dps-dpsapex}, and we found that now DPS works on a batch size of 4*4 without an OOM error. This phenomenon is probably because Apex uses a mixture of \texttt{fp16} and \texttt{fp32} data to reduce the data amount stored on GPU and thus avoid GPU OOM effectively. A detailed explanation of how this reduces memory usage can be found in Appendix \ref{sec:apex}. It's also interesting to note that this solution further decreases the training time to half of the original DPS implementation, as discussed in Section \ref{sec:explain-apex}. Another conclusion gained from Figure \ref{fig:dps-dpsapex} is that the optimization options (\texttt{O1} and \texttt{O2}) offered by Apex do not have a big difference on GPT-2.

Inspired by the performance improvement brought by Apex \texttt{fp16}, we conducted the same approach with Horovod, with the results shown in Figure \ref{fig:hvd-hvdapex}. It is clear that Horovod has larger performance improvement than DPS, statistically by a factor of 2. It's also observed that Horovod-Apex can even compete with DPS-Apex now, and at some point ($<3m$), it even outperforms DPS-Apex.

We now analyze a 2-node configuration running Horovod-Apex. According to Figure \ref{fig:hvdapex-4-8}, we can hardly see obvious differences of the curves under 4 nodes and 8 nodes. However, if we pick the same total batch size, a batch size of 2*8 (with 8 GPUs) on 2 nodes is even slower than 4*4 (with 4 GPUs) on one node.

This is probably because the speed of communication among multiple nodes is slow, and the communication payload increases. As discussed in Section \ref{sec:data}, unlike communication of GPUs, communication of nodes costs much more time because of the network protocol. The communication payload increases because the reduction of the batch size on one GPU requires a more frequent synchronization. As the data length of the gradient does not change for each synchronization, the total communication workload is incremented. These two factors cause the converging speed of the training process to be even slower.

\begin{figure}[htp!]
	\begin{center}
  		\includegraphics[width=1.0\linewidth]{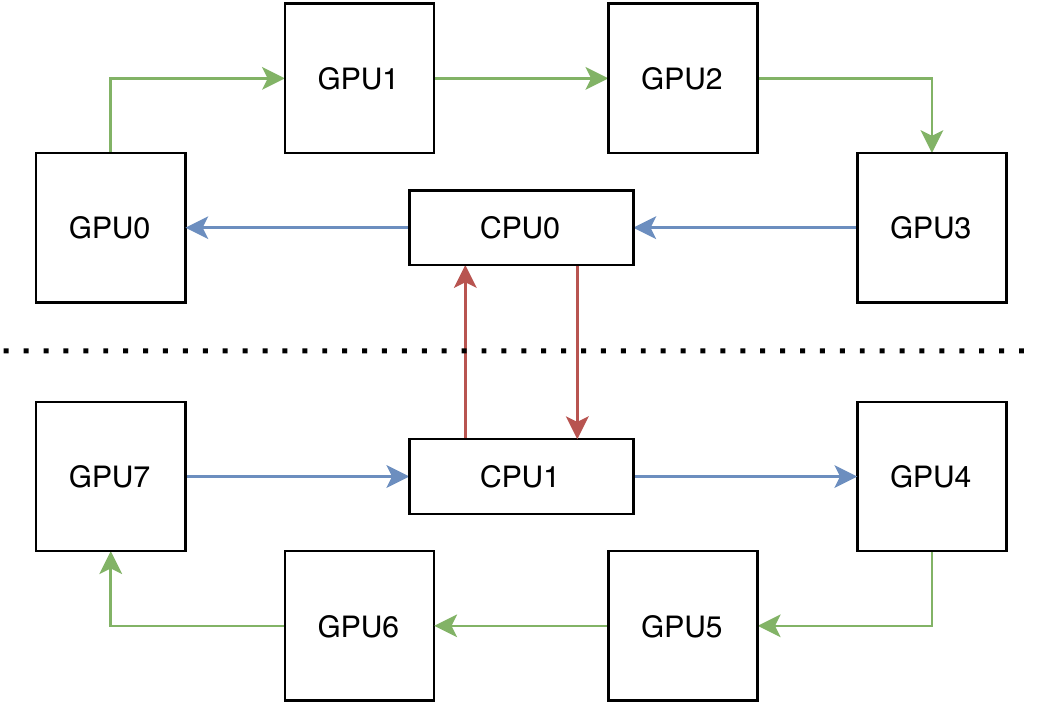}
	\end{center}
   	\caption{Gradient communication routine in 2-node configuration Horovod (Ring-Allreduce) architecture. Blue line indicates inter-GPU communication, blue line indicates GPU-CPU communication, and red line indicates inter-node communication, which becomes a performance bottle-neck.}
	\label{fig:bottleneck2node}
\end{figure}

This conclusion can be well supported considering the performance of Horovod-Apex under the batch size of 4*8. As illustrated, different from the relationship in Figure \ref{fig:dps-hvd-2-4}, the training process under a larger batch size is accelerated, making it faster than both batch sizes of 4*4 and 2*8 on Horovod-Apex, which is probably because the performance increase in GPU is greater than the performance decrease due to inter-node communication. However, we still see the performance increase is much lower than the expected value (double) as the inter-node communication forms a bottleneck in the network, as illustrated in Figure \ref{fig:bottleneck2node}.

The aggregation of the results is illustrated in Figure \ref{fig:all} for reference.\footnote{The red line (Horovod-Apex) is cut off in the end because it experiences a gradient overflow issue from time to time after that point. But according to official explanations, it does not influence performance.} According to the results, it is clear that Horovod-Apex, DPS, and DPS-Apex outperform other methods by a large margin when only one node is reachable, and DPS-Apex is performing the best on one node. 

\section{Conclusion}

In conclusion, in the context of \texttt{PyTorch}, under the limitation of one node, it's recommended to use DPS-Apex to gain the best performance. However, DPS requires a higher GPU memory and is prone to out-of-memory errors. On the contrary, it is more robust for Horovod with the same batch size because of its memory cleaning and workload balancing capabilities. Although Horovod is a bit slower, its robustness provides a very good experience for the training process. With the help of Apex \texttt{fp16}, the performance of Horovod can be lifted up to merely slightly lower than DPS-Apex.

Under the condition of multiple nodes, the operating difficulty of DPS is much larger than Horovod, as it requires running the command on every node concerned. DPS is also not scalable on multiple nodes, because it still requires one master GPU for the AllReduce process. Horovod is feasible for multiple nodes because it can be launched using a single command \texttt{horovodrun} and every node will automatically call \texttt{mpirun} to conduct the task on their GPUs, and it uses the Ring AllReduce for synchronization, which is fully balanced for each GPU worker.

\section{Future Work}
The future work of this research will be focused on two topics:

\begin{itemize}
	\item Quantitative Results for Data Parallelism on Tensorflow. Although we've completed data parallelization on PyTorch and analyzed the qualitative expectations of this on Tensorflow, we have not implemented the model on Tensorflow, so we lack a comparison between PyTorch and Tensorflow.
	\item Quantitative Results for Model Parallelism on PyTorch. We've completed the work to do data parallelism, but we still lack first-hand evidence to show the power of model parallelism.
\end{itemize}

We're going to work on this topic so as to achieve these two future goals.

\begin{acks}
I would like to sincerely thank Prof. Volodymyr Kindratenko for guiding this research and Nuohan Yang for reviewing this manuscript, and Dr. Dawei Mu for providing technical support with the HAL system. This work utilizes resources supported by the National Science Foundation's Major Research Instrumentation program, grant \#1725729, as well as the University of Illinois at Urbana-Champaign.
\end{acks}

\bibliographystyle{ACM-Reference-Format}
\bibliography{text.bib}

\clearpage

\appendix

\section{The SLURM Queueing System} \label{sec:slurm}

Proposed in 2003, the SLURM (Simple Linux Utility for Resource Management) task manager is a framework for multi-user hierarchical control \cite{yoo2003slurm}, which is now widely accepted by companies and academia. As super-computers are often used by multiple users in an organization, Linux needs to know how to schedule all the tasks received and make a relatively rule for each user to share computational resources.
	
When the code and data are in place, the user is required to write a script to launch the task, including which nodes to use, how many GPUs are needed, and the maximum time for executing the script. The script is called by \texttt{sbatch <script>}, so it becomes the entry point of the whole program. Essentially, it's just a \texttt{bash} script indicating resources required of the task, and what the task does. The script will be executed in a blocking way, so the second command in the script must wait until the first command is completed and returned.
	
As SLURM assigns the user tasks randomly, it's hard for the user to know which node he is allocating before he launches the script. This problem can be solved by requesting a reservation on SLURM.

\section{Multi-Head Attention Layer} \label{sec:transformer}

The multi-head attention layer is the crucial concept proposed by the transformer architecture to parallelize temporal data modeling. As this layer is crucial in the GPT-2 model, we give a mathematical introduction here for readers to refer to.

\subsection{Variances Of Recurrent Neural Network}

\begin{figure}[htp!]
	\begin{center}
  		\includegraphics[width=0.7\linewidth]{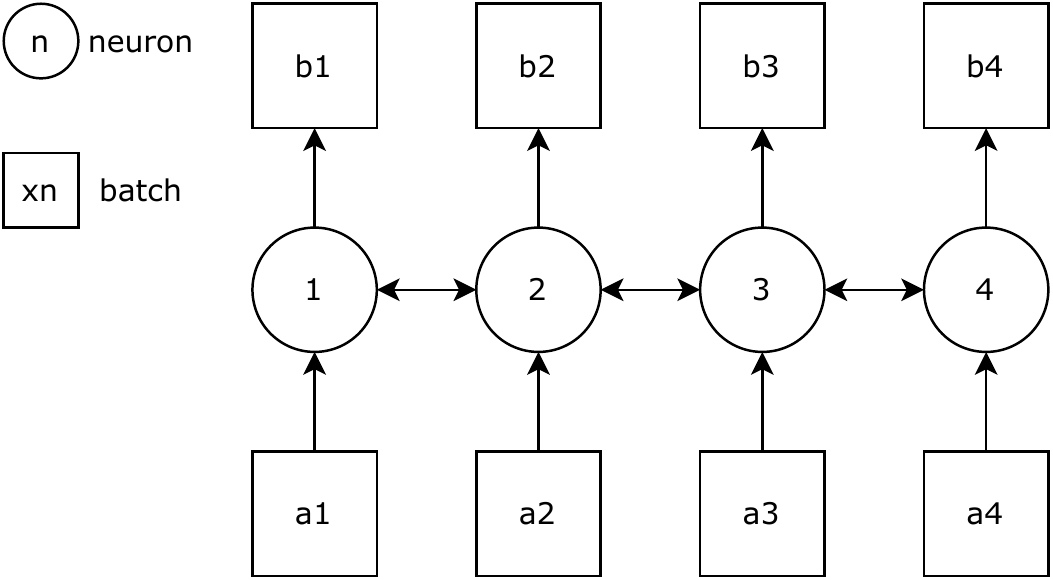}
	\end{center}
   	\caption{The architecture of Bidirectional RNN.}
	\label{fig:rnn}
\end{figure}

Before the transformer was proposed, the most popular sequential models were all done by the Recurrent Neural Network (RNN). The basic ideology of each RNN layer is to input a sequence of vectors (words) and output a sequence of vectors with the same position and length. In other words, each RNN layer turns a sentence into another sentence with the same length. Because RNN cannot get future information, bidirectional RNN is proposed, as shown in Figure \ref{fig:rnn}. 

\begin{figure}[htp!]
	\begin{center}
  		\includegraphics[width=0.8\linewidth]{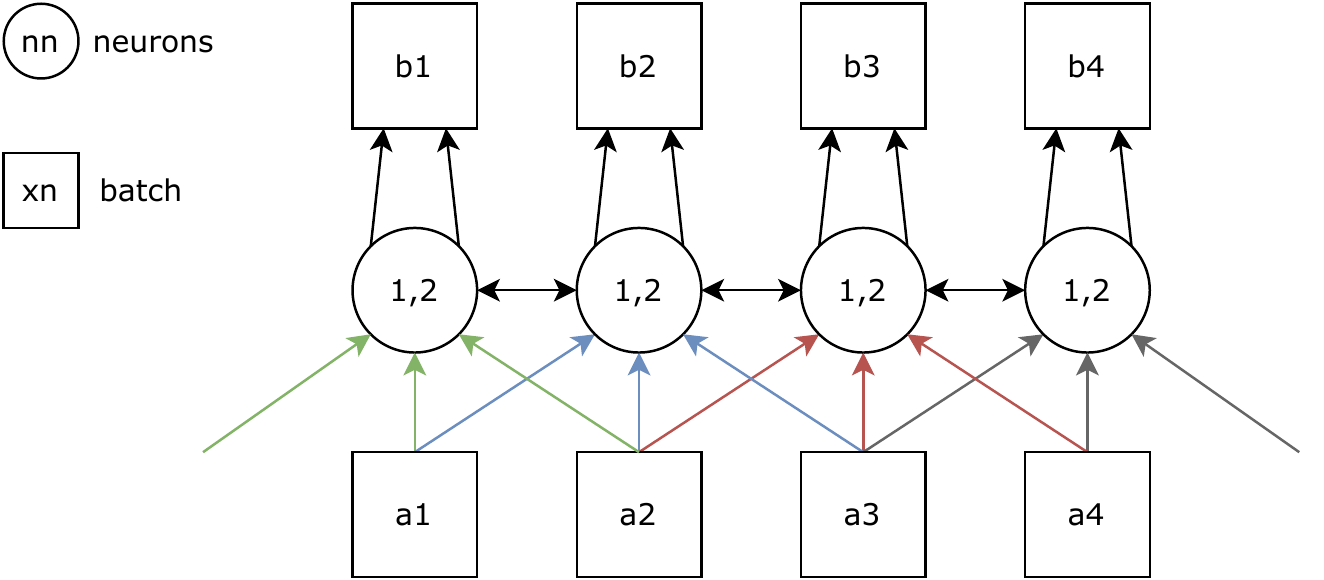}
	\end{center}
   	\caption{The architecture of Convolution RNN.}
	\label{fig:crnn}
\end{figure}

As indicated by its name, the communication among the neurons is bidirectional, which means a neuron can get information from both sides of a word in the sentence. However, RNN cannot be parallelized. For example, to get the result $b_4$, we have to first calculate $b_1$ to $b_3$. To solve this problem, the idea of \textbf{Convolutional RNN} is proposed: CNN be used to replace RNN, as shown in Figure \ref{fig:crnn}. 

In the figure, each CNN neuron takes 3 inputs: the current word, the last word, and the next word. Since the output of each neuron is only a scalar, we use multiple neurons to represent the whole vector. For example, if the input vector has a dimension of 2, or $len(a_i)=2$, we need 2 neurons to produce 2 scalars to construct the vector. This architecture is much easier for parallelization. 

However, CRNN is still not satisfying, because each CNN neuron can only consider a very limited amount of information. In the last example, we can only detect 3 vectors for one CNN neuron. We call the number of detectable vectors Receptive Field. This leads to much worse performance for CNN compared to RNN, as RNN makes use of all the information from previous neurons or even the whole sentence. 

\begin{figure}[htp!]
	\begin{center}
  		\includegraphics[width=0.6\linewidth]{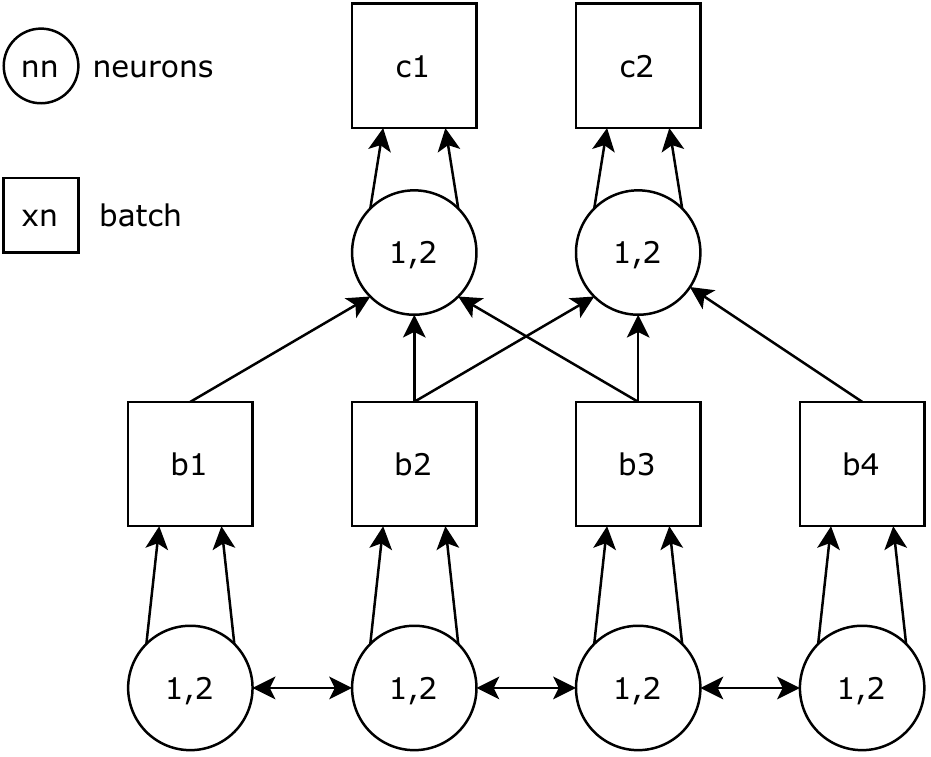}
	\end{center}
   	\caption{The architecture of the Hierarchical CRNN.}
	\label{fig:crnn-hierarchy}
\end{figure}

An attempt to solve this problem is a hierarchical design of the CNN layers, as shown in Figure \ref{fig:crnn-hierarchy}. However, this model still does not improve the receptive field for the first neuron.

\subsection{The Self-Attention Layer} \label{sec:self-attention}

To solve this problem completely, the idea of the self-attention layer was introduced. The task of the self-attention layer is exactly the same as RNN - it takes a sentence as input, and outputs a sentence of the same size. The difference is that all the words are calculated and output at the same time, so they can be paralleled and easier for GPUs to process. Simply put, the self-attention layer is a paralleled RNN.

\begin{figure}[htp!]
	\begin{center}
  		\includegraphics[width=0.8\linewidth]{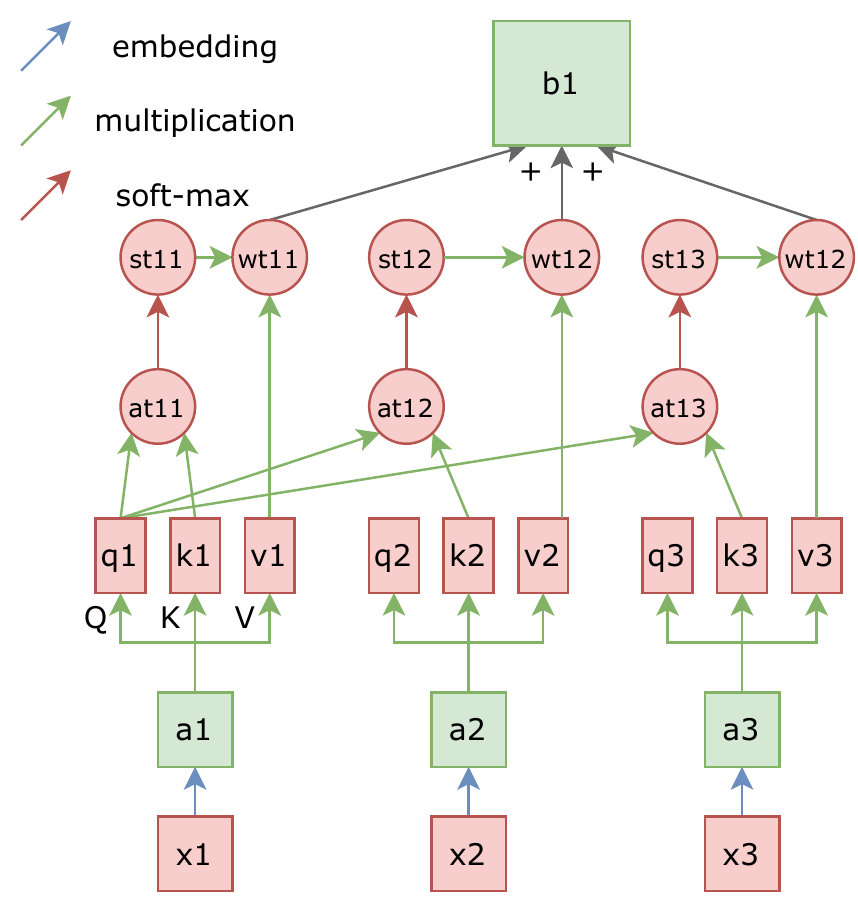}
	\end{center}
   	\caption{The architecture of a self-attention layer.}
	\label{fig:self-attention}
\end{figure}

An example of the architecture of the self-attention layer is shown in Figure \ref{fig:self-attention}. The input is three words labeled $x_i$, and each of them goes through an embedding function to get its word vector $a_i$. Then the model defines 3 matrices, which are called the attention heads: the Q matrix for query, the K matrix for key, and the V matrix for weighted word vector. Practically, the query is used to match all keys to produce the attention vector $\alpha_{ij}$, which is represented as "at" in the figure. We can represent the calculations in Formula \ref{equ:self-attention}.

\begin{align} \label{equ:self-attention}
	\alpha_{ij}&=\frac{q_i\cdot k_j}{\sqrt{d}} \\
	q_i &= Qa_i \\
	k_i &= Ka_i \\
	v_i &= Va_i 
\end{align}

The $\alpha_i$ attention vectors are then normalized by passing through the soft-max layer and become the standardized attention vectors $s_i$. This vector does a dot-product with the weighted word vector $v_j$ and gains the weight vector $w_j$ for this word. This is why $v_i$ is called the weighted word vector - it is directly extracted from the embedded word and be summed up with the other vector, thus gaining the weight of this word. The word vector can represent the information stored in the word very well.

At last, we sum up the three weight vectors and get the final result $b_1$. Repeating this process 3 times, we can get the corresponding output vector $b_2, b_3$ for $a_2, a_3$ too, and thus outputting exactly the same form of data like RNN - one $b_i$ for each $a_i$.

Looking back at the gained vector $b_i$, we surprisedly find that it has already taken all information in the input sentence $A$. Consider the case we have a sentence of length $l>3$, the parallization will be much more apparent and thus $b_1$ has a larger receptive field with a longer sentence. 

However, for the naive model shown above, it's still not obvious how everything can be parallelized. For example, we have to get $s_{11}$ to get $w_{11}$. In Section \ref{sec:par-self-attention}, we dicuss how to parallelize the calculation using linear algebra (matrix).

\subsection{Parallization Of The Self-Attention Layer} \label{sec:par-self-attention}

Firstly, we notice that the transformer matrices $Q, K, V$ (we now represent them as $W^Q, W^K, W^V$ as they're essentially weights) for each word are the same. Thus, we can use one matrix to calculate all the results in one-shot using the formula $[q_1q_2q_3q_4]=W^Q[a_1a_2a_3a_4]$, where $[q_1q_2q_3q_4]$ is the matrix constructed by stacking the four query vectors. We apply the same strategy to $W^K$ and $W_V$ for parallelization.

\begin{align} \label{equ:par-self-attention-1}
	Q = W^QA, K&=W^KA, V=W^VA\\
	Q = [q_i]_1^l, K &= [k_i]_1^l, V = [v_i]_1^l \\
	A &= [a_i]_1^l 
\end{align}

Secondly, we observe that the calculation of the attention vectors $\alpha_i$ can be parallelized because they're all using the same query vector $q_i$. Thus, we construct the attention matrix $\mathcal{A}$ as shown below.

\begin{align} \label{equ:par-self-attention-2}
	\left[ \begin{matrix}\alpha_{11} \\ \alpha_{12} \\ \alpha_{13} \end{matrix} \right]  =\left[ \begin{matrix}k_{1}\\ k_{2}\\ k_{3}\end{matrix} \right]  \cdot q_{1}
\end{align}

We can apply the same process to $q_2$ and $q_3$, and we find that the $Q$ matrix can be used to solve the whole problem.

\begin{align} \label{equ:par-self-attention-3}
	\left[ \begin{matrix}\alpha_{11} &\alpha_{21} &\alpha_{31} \\ \alpha_{12} &\alpha_{22} &\alpha_{32} \\ \alpha_{31} &\alpha_{32} &\alpha_{33} \end{matrix} \right]  =\left[ \begin{matrix}k_{1}\\ k_{2}\\ k_{3}\end{matrix} \right]  \cdot \left[ \begin{matrix}q_{1}&q_{2}&q_{3}\end{matrix} \right]  
\end{align}

If we generalize the matrix, we get

\begin{align} \label{equ:par-self-attention-4}
	\mathcal{A}=(\left[ k_{i}\right]^{l}_{1}  )^{T}\cdot \left[ q_{i}\right]^{l}_{1}=K^TQ
\end{align}

Then we apply the softmax function and get 

\begin{align} \label{equ:par-self-attention-5}
	\hat{\mathcal{A}} = Sm(\mathcal{A})
\end{align}

At last, we can parallelize the last step as well. Taking our example as the starting point, we have

\begin{align} \label{equ:par-self-attention-6}
	b_{1}&=[v_{1}v_{2}v_{3}]\cdot \left[ \begin{matrix}\hat{\alpha }_{1} \\ \hat{\alpha }_{2} \\ \hat{\alpha }_{3} \end{matrix} \right] \\
	b_{1}b_{2}b_{3}&=[v_{1}v_{2}v_{3}]\cdot \left[ \begin{matrix}\hat{\alpha }_{11} &\hat{\alpha }_{21} &\hat{\alpha }_{31} \\ \hat{\alpha }_{12} &\hat{\alpha }_{22} &\hat{\alpha }_{32} \\ \hat{\alpha }_{13} &\hat{\alpha }_{23} &\hat{\alpha }_{33} \end{matrix} \right]  
\end{align}

To put in matrix, we have the simplified form

\begin{align} \label{equ:par-self-attention-7}
	B=V\cdot \hat{\mathcal{A}}
\end{align}

Now we make a conclusion of the task done by the self-attention layer using matrix. First, we get the intermediate matrices - the query matrix, the key matrix, and the vector matrix. Then we get the attention matrix using the key matrix and the queue matrix. After that we apply the soft-max function on the attention matrix to get the standardized attention matrix. At last, we use the vector matrix and the standardized attention matrix to get the output matrix $B$. The whole process is shown below.

\begin{align} \label{equ:par-self-attention-8}
	Q = W^QA, K&=W^KA, V=W^VA \\
	\mathcal{A} &= K^TQ \\
	\hat{\mathcal{A}} &= Sm(\mathcal{A}) \\
	B &= V\hat{\mathcal{A}}
\end{align}

\subsection{Multi-head Attention}

\begin{figure}[htp!]
	\begin{center}
  		\includegraphics[width=0.8\linewidth]{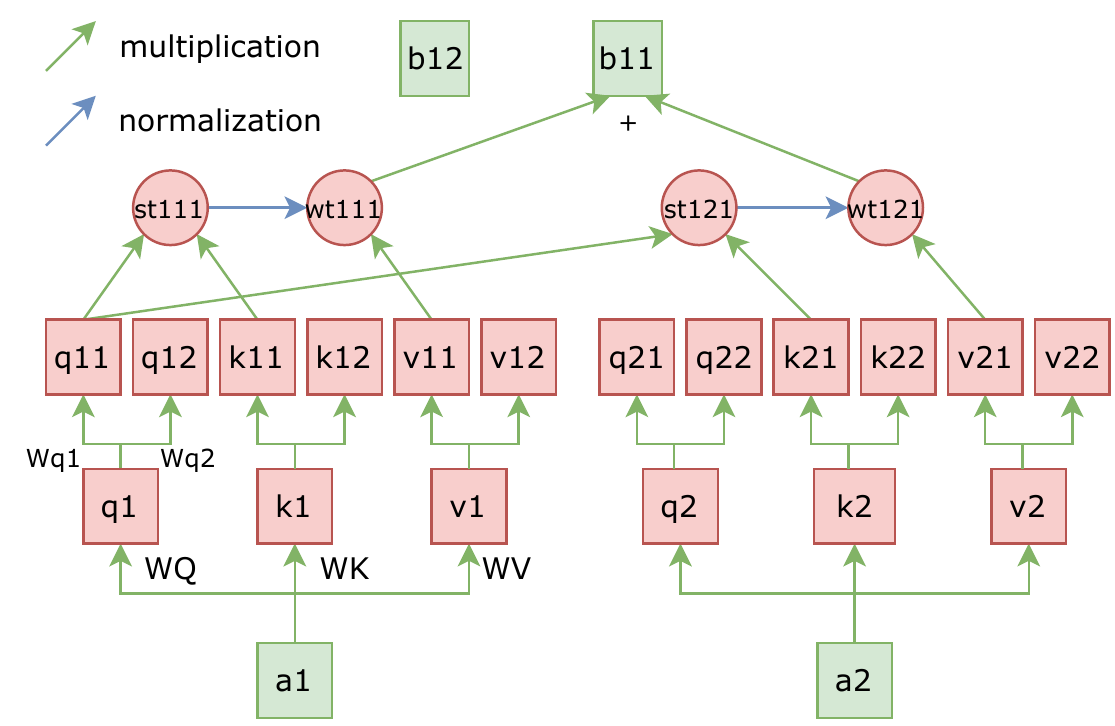}
	\end{center}
   	\caption{The architecture of a multi-head attention layer.}
	\label{fig:multihead-attention}
\end{figure}

The multi-head attention splits the attention head up to multiple smaller heads, and the architecture is shown in Figure \ref{fig:multihead-attention}. Using the multi-head attention layer, the dimension of the features is further expanded to a factor of the number of attention heads. 

For each attention head query $q_i$, we use matrices to further split it. In the example in the figure, we have 2 multihead attention matrices. Taking the query matrix as example, the formula is shown below.

\begin{align} \label{equ:par-self-attention-9}
	q_{ij} = q_i W_{qj}
\end{align}

This process happens exactly the same for vectors $k$ and $v$. In the example above, now we get all the attention heads $q_{im}, k_{im}, v_{im}$, where $m$ represents the number of multi-head heads. Then we use the splitted attention head $q_{11}$ to calculate each attention with $k_{11}, k_{21}$, and thus get $\alpha_{111}, \alpha_{121}$ and $w_{111}, w_{121}$. Finally, we dot product the two weight vectors and get $b_{11}$. After getting all the $b_{im}$ results, we can stack them and multiply by an output matrix $W_b$, as shown below.

\begin{align} \label{equ:par-self-attention-10}
	b_{i}=W_{b}\left[ \begin{matrix}b_{i1}\\ ...\\ b_{im}\end{matrix} \right]  
\end{align}

This dimension expansion strategy is used to detect more information in the system. For example, if we split one head into 2, with each head multiplied by a weight matrix, we can train the parameters so that one split head is used for detecting the information of the neighbor words, and the other is used for detecting the global information. 

\subsection{Positional Encoding}

As mentioned in Section \ref{sec:self-attention}, the self-attention layer takes all information in the sentence to train one word. Thus, this word has no information about where the words it's learning from are in the sentence. This can become a serious problem, as failure to locate words makes the transformer lose the ability that RNN has. To overcome it, the Positional Encoding technique is introduced, which adds a position vector $e_i$ to the embedded word $a_i$.

This process has a prototype of concatenation. Suppose we have a one-hot positional vector $p_i$, in which the only number $1$ indicates the position of the word in a sentence, then we have the formula:

\begin{align} \label{equ:par-self-attention-11}
	\left[ \begin{matrix}W_{A}&W_{P}\end{matrix} \right]  \left[ \begin{matrix}x_{i}\\ p_{i}\end{matrix} \right]  =W_{A}\cdot x_{i}+W_{P}\cdot p_{i}=a_{i}+e_{i}
\end{align}
		
\section{Insight of GPU Memory Usage} \label{sec:rules-gpu-mem}

In our research, we encountered lots of GPU Out Of Memory (OOM) issues, and we managed to propose several possible solutions to it by investigating the essence of the error. In Appendix \ref{sec:rules-gpu-mem}, we're going to analyze the GPU usage in the training process, and in Appendix \ref{sec:reduce-gpu-mem} we're going to solve the OOM errors and discuss why the errors were encountered in our research.

\subsection{The CUDA GPU Memory} \label{sec:cuda-mem}

GPU memory is different from CPU memory because GPU memory is exclusively used by the GPU streaming multiprocessors on the card, while CPU memory (which is often called the general memory or simply ``memory") is used by a series of devices including the CPU itself. The bandwidth of GPU memory (GDDR6) is 3 times faster than the general memory (GDDR4) due to a great requirement of high-performance computing, and thus it's not recommended to train a deep neural network on a CPU instead of a GPU. 

However, as CPU memory is commercialized and can be installed by the user himself, the size of CPU memory is often much larger than the GPU memory. Thus, under extreme circumstances\footnote{Extreme circumstance means cases where no multi-GPU solution is possible, a large amount of memory is required to start the training (more than one GPU can satisfy), and cannot solve by reducing the model size.}, it's still a valid way to change GPU memory to CPU memory to get the OOM issue solved, as will be discussed in Appendix \ref{sec:cpu-memory}.

\subsection{Estimating the Mamory Use of Models} \label{sec:model-mem}

Before solving the OOM issue, we need to first estimate which part of the model causes the usage of CUDA memory \cite{Chenyun_2018}. Basically, it's divided into three parts: parameters of the model, parameters of the optimizer for the model, and I/O for each layer in the model.

\subsubsection{Parameters of the Layer ($p_l$)} \label{sec:param-model}

The number of parameters of the model is simply the sum of the weight and bias in all layers in the model, and doesn't depend on the size of input data. In detail, the model is loaded after the command \texttt{model = MyGreatModel(). cuda()} or \texttt{model.to(cuda:\{i\})}, and can be examined using the command \texttt{model.parameters()}. It should also be noted that only some layers have weights (like CNN layer, RNN layer, FCN layer, BatchNorm layer, and Embedding Layer), while the others do not (like activation layer, dropout layer, and pooling layer) \cite{gao2020estimating}. Table \ref{tab:layer-parameter} is a summary of the layers with their layer weights amount.

\begin{center}
\begin{table}[htp!]
\centering
\small
\setlength\tabcolsep{2.3pt}
\begin{tabular}{c|cc|c}
\hline
Layer & Layer Input & Layer Output & Weights Amount $p_l$\\
\hline
Linear & $m$ & $n$ & $m \cdot n$ \\
Conv2d & $c_{i}, k$ & $c_{o}$ & $c_{i}\cdot c_{o}\cdot k^2$ \\
BatchNorm2d & $n, c, h, w$ & $n, c, h, w$ & $2\cdot n$ \\
Embedding & $n, w$ & $n, w, h$ & $n \cdot w$ \\
\hline
\end{tabular}
\caption{Parameter number in memory for one layer in PyTorch, originally proposed by \cite{xiaoyao_2021}.}
\label{tab:layer-parameter}
\end{table}
\end{center}

Note that the number of parameters is not the same as the memory usage in memory, as each parameter has a data size. For example, \texttt{float32} will multiply the number argument by 4 to get the memory usage, \texttt{double} will multiply it by 8, and so on.

\subsubsection{Parameters of the Layer After Optimization ($p_r$)} \label{sec:param-opt}

The parameter of the optimizer is the parameter generalized when back-propagating in the optimization state. In other words, the parameter of the optimizer is the gradient ($\Delta F(w)$) and momentum ($v$).

As different optimizers utilize different optimization technologies, the actual amount of the parameters of the optimizer is also variable. For \texttt{SGD}, it takes approximately the same parameters as the model, which means that the parameter amount will be doubled after this optimizer. For \texttt{Adam}, it takes approximately 3 times as much as the model parameter, including the size of SGD, and double the size of momentum information, which means that the total parameters will be 4-fold after optimization. For reference, the memory storage of the optimizers is shown in Table \ref{tab:opter-parameter} \cite{bakhoda2009analyzing}.

\begin{center}
\begin{table}[htp!]
\centering
\small
\setlength\tabcolsep{2.3pt}
\begin{tabular}{c|ccc|cc}
\hline
Layer Optimizer & Layer & Gradient & Momentum & Result $p_r$ & Factor $n$\\
\hline
SGD & $p_l$ & $p_l$ & $0$ & $2p_l$ & $2$ \\
SGD-Momentum & $p_l$ & $p_l$ & $p_l$ & $3p_l$ & $3$ \\
Adam & $p_l$ & $p_l$ & $2p_l$ & $4p_l$ & $4$ \\
\hline
\end{tabular}
\caption{Parameter number in memory for one layer after applying the optimizer in PyTorch, originally proposed by \cite{Chenyun_2018}.}
\label{tab:opter-parameter}
\end{table}
\end{center}

In PyTorch, the optimizer is not layer-specific, as each optimizer applies to all layers in a model. Thus, we first calculate the model parameter without the optimizer and then multiply it by the optimizer factor. If we represent $p_r^{l_i}$ as the parameter in layer $i$, $l_i$ as the layer number, and $p_m$ as the parameter in the whole model, then the parameter number in the whole model can be calculated by Formula \ref{equ:layer-to-model}, as shown below.

\begin{align} \label{equ:layer-to-model}
	p_m &= \sum^{n}_{i=0}p_l^{l_i} \cdot n
\end{align}
	
\subsection{Layer Output and Model Input} \label{sec:io-layer}

Except for memory for each layer, we also need memory for data produced by each layer. Thus, if we represent $p_o^{l_i}$ as the sum of all the outputs of layer $i$, the total memory usage for storing all the parameters for each layer output can be calculated by Formula \ref{equ:output-layer-to-model}. This formula holds because the output of each layer is the input of the next layer - we don't need to add the layer inputs in the formula.

\begin{align} \label{equ:output-layer-to-model}
	p_o &= \sum^{n}_{i=0}p_o^{l_i}
\end{align}

\begin{figure}[htp!]
	\begin{center}
  		\includegraphics[width=1.0\linewidth]{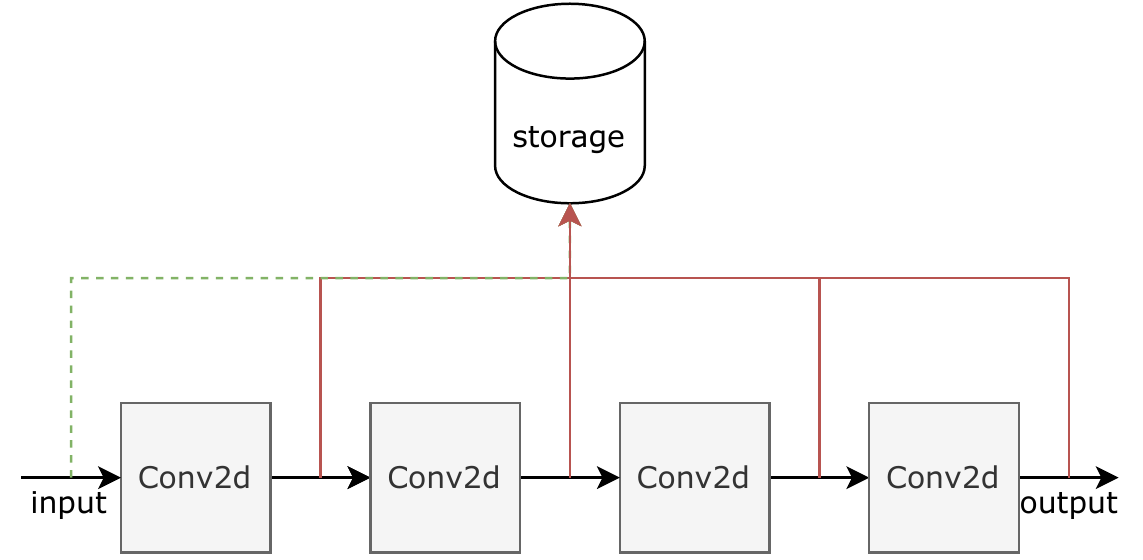}
	\end{center}
   	\caption{The illustration of storage of output of each layer $p_o$ (the read lines) and the storage of input of the model $p_b$ (the dashed green line). The latter one is usually ignored because we typically don't calculate it's gradient, and it's usually split into pieces so it does not occupy too much storage.}
	\label{fig:data-layer-storage}
\end{figure}

The size of input data for the model can usually be omitted, because we always use an iterator to split the data into batches - a 5-GB large dataset can be split up into a large amount of 1-MB data, which is handled by the data loader. The illustration of the layer output and model input contributing to memory storage is in Figure \ref{fig:data-layer-storage}.

For the question about $p_o$ in the case that the input data is large: although the input data of the model $p_b$ is also multiplied by a large number, causing a large memory usage, it is still relatively small than $p_o$.

\subsection{Concluded Formula}

Taking all 3 ingredients into consideration, we have the CUDA memory calculation Formula \ref{equ:mem_calc}, as shown below.

\begin{align} \label{equ:mem_calc}
	\mathcal{M} &= p_m \cdot n + b \cdot p_o + p_b \\
	p_m * n &= \sum^{n}_{i=0}p_r^{l_i}
\end{align}

Here $\mathcal{M}$ is the CUDA memory usage, $p_m$ is the parameter of the model, $p_o$ is the sum of parameters of the output for each layer, $p_b$ is the size of the input data for the whole model, $b$ is the batch size, and $n$ is an integer usually between 2 and 6, which is subjective to the type of optimizers, like 2 for SGD and 4 for Adam. Note that the last item ($p_b$) can usually be ignored as discussed in Section \ref{sec:io-layer}. 
	
\section{Reducing GPU Memory Usage} \label{sec:reduce-gpu-mem}

In this appendix, we illustrated the practical ways to reduce memory usage and the reason for the unexpected OOM issues in our experiments, as illustrated in Section \ref{sec:graphical}. In this part, we frequently refer to Formula \ref{equ:mem_calc} so it's recommended to go through Appendix \ref{sec:rules-gpu-mem} first. 
	
\subsection{Change Data Length using Apex} \label{sec:apex}

The most efficient strategy, also the easiest one, is to shift the processed data to another type with a shorter length. For example, if the data is now  floating-point-32, consider shifting it to floating-point-16 for all data, including the model and the input data. This approach decreases a bit of the precision, but it decreases the usage of the memory to approximately a half  \cite{Laosong_2021}.

This can be proved by Formula \ref{equ:mem_calc}. Because all the parameter sizes ($p_m, p_o, p_b$) are decreased by a factor of 2, the whole usage $\mathcal{M}$ is also decreased by a factor of 2. 

However, changing data types manually is time-consuming and bug-prone, and can also decrease the precision non-trivially. Thus, PyTorch developed a very powerful library utilizing NVIDIA Tensor Core called Apex. The Apex library takes floating-point-32 as original code but utilizes in-library optimizations to the data using a mixed distribution of floating-point-16 and floating-point-32 to decrease the usage of data to a factor of approximately 2. This strategy is proved to lose only a trivial precision when training and requires adding only a few lines into the original code with floating-point-32 data.
	
However, mixed-precision training using \texttt{fp16} (half precision) is now only available for GPUs supporting NVIDIA Tensor Core. Fortunately, the \textit{HAL} system is built using V100 GPUs based on the Volta architecture, which has the built-in Tensor Core, so it's safe to use in our experiments.

The detailed code snippet for utilizing the Apex library can be found on the official website at \url{https://github.com/NVIDIA/apex}.

\subsection{Minimize Batch Size} \label{sec:minimize-batch-size}

Minimizing batch size is another crucial method to reduce $\mathcal{M}$. According to Formula \ref{equ:mem_calc}, when $b$ is halved, the second item is halved. Thus, when the first item is small,\footnote{The size of the layers are usually much smaller than the output data size of the layers.} this is very useful for reducing the memory usage. Note that reducing both development batch size and testing batch size can contribute to a smaller $\mathcal{M}$ \cite{xiaoyao_2021}.

\begin{figure}[htp!]
	\begin{center}
  		\includegraphics[width=1.0\linewidth]{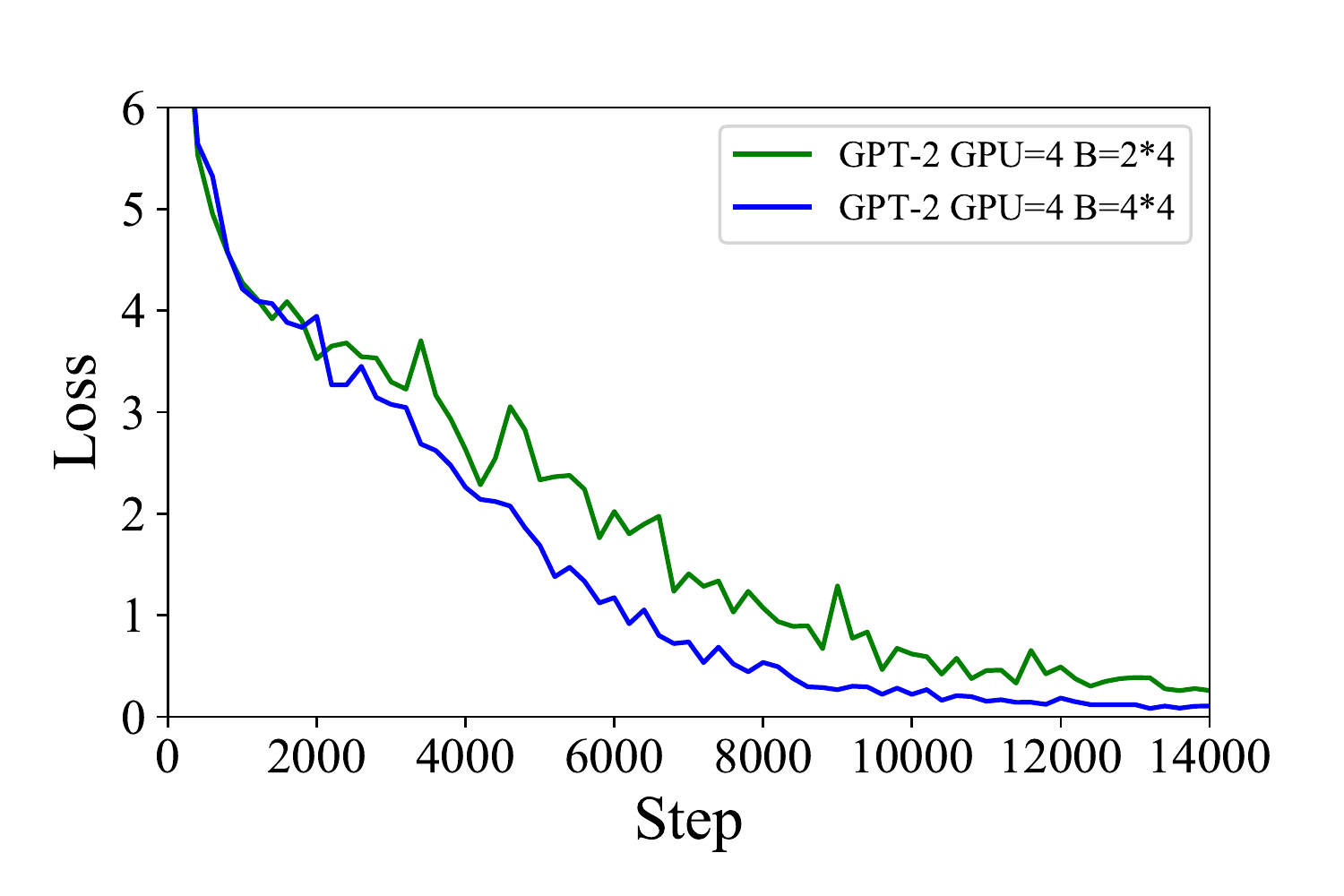}
	\end{center}
   	\caption{The loss curve of running GPT-2 under 4 GPUs with batch size 2*4 and 4*4.}
	\label{fig:hvd-2-4}
\end{figure}

However, small batch size is very likely to exert a performance loss of the convergence of the loss curve. For example, in our GPT-2 training process, if the batch size becomes 1 or 2, the model will not be sufficiently fed for each step and thus create a tendency of divergence in the loss curve. We've done several experiments, with the results shown in Figure \ref{fig:hvd-2-4}. The results show that a smaller batch size tends to make the loss curve not as smooth and prone to diverge. Even if we observe a normal loss curve, a small batch size may also lead to results that are not meaningful.
	
Thus, researchers should double-check the influence of reducing batch size in the model they're investigating, as the conclusion varies from model to model.
	
\subsection{Reduce Input Sequence Length} \label{sec:minimize-seq-length}

Reducing the input sequence length is another popular way to tackle the OOM issue. Note that this method is more likely to be used in a sequential model that is order-sensitive, such as data used in RNN models. In these kinds of training, the sequence length of the data is linear to $p_o$, thus a halved sequence length has the same effect as halved batch size $b$, leading to approximately the same memory saving as discussed in Section \ref{sec:minimize-batch-size}.
	
\subsection{Free Unnecessary Variables} \label{sec:free-vars}

    \begin{figure*}
        \centering
        \begin{subfigure}[b]{0.475\textwidth}
            \centering
            \includegraphics[width=\textwidth]{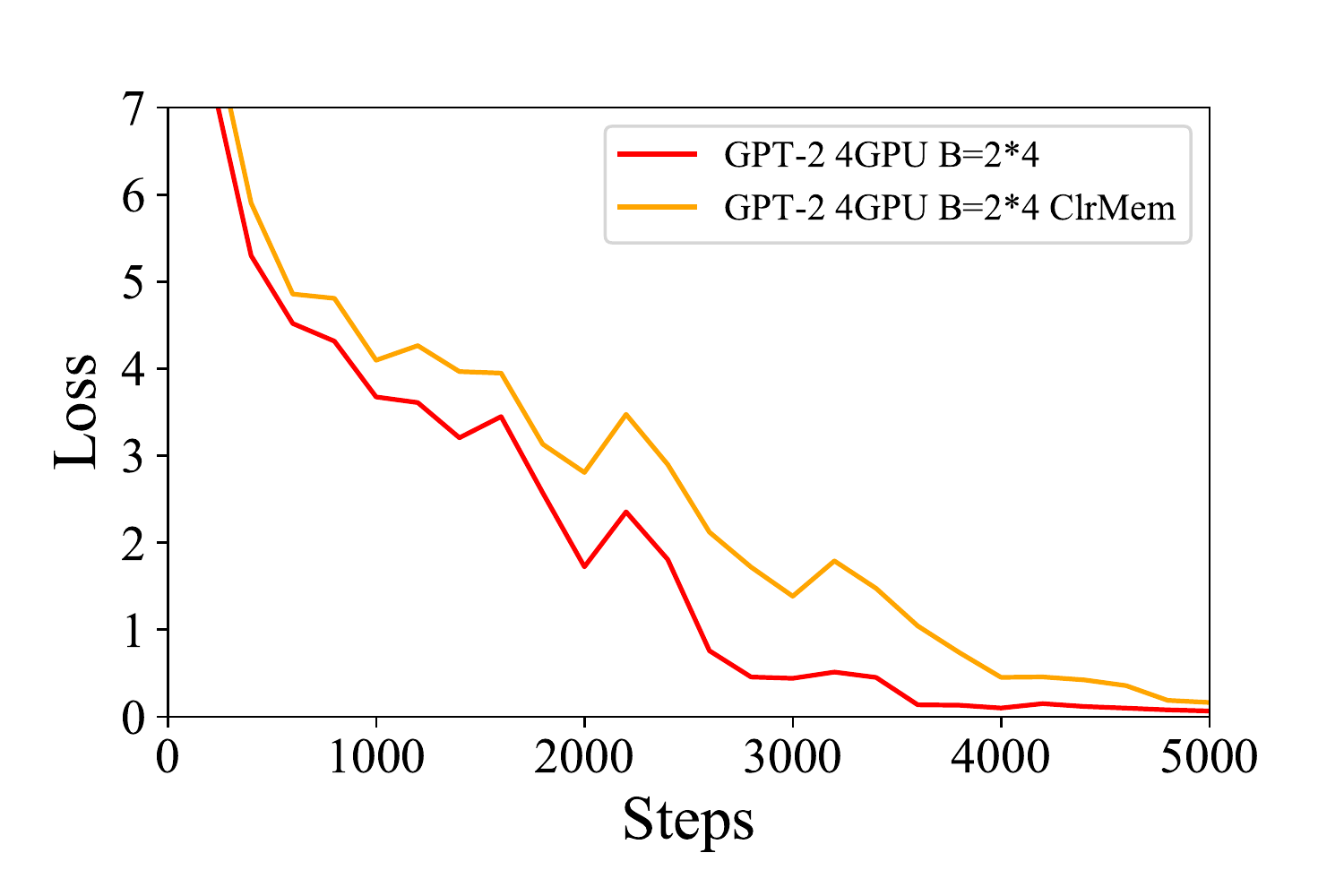}
            \caption[]%
            {{\small The loss-step curve of running GPT-2 under 4 GPUs with batch size 2*4 with memory clear and without memory clear.}}    
            \label{fig:dps-step-clrmem}
        \end{subfigure}
        \hfill
        \begin{subfigure}[b]{0.475\textwidth}  
            \centering 
            \includegraphics[width=\textwidth]{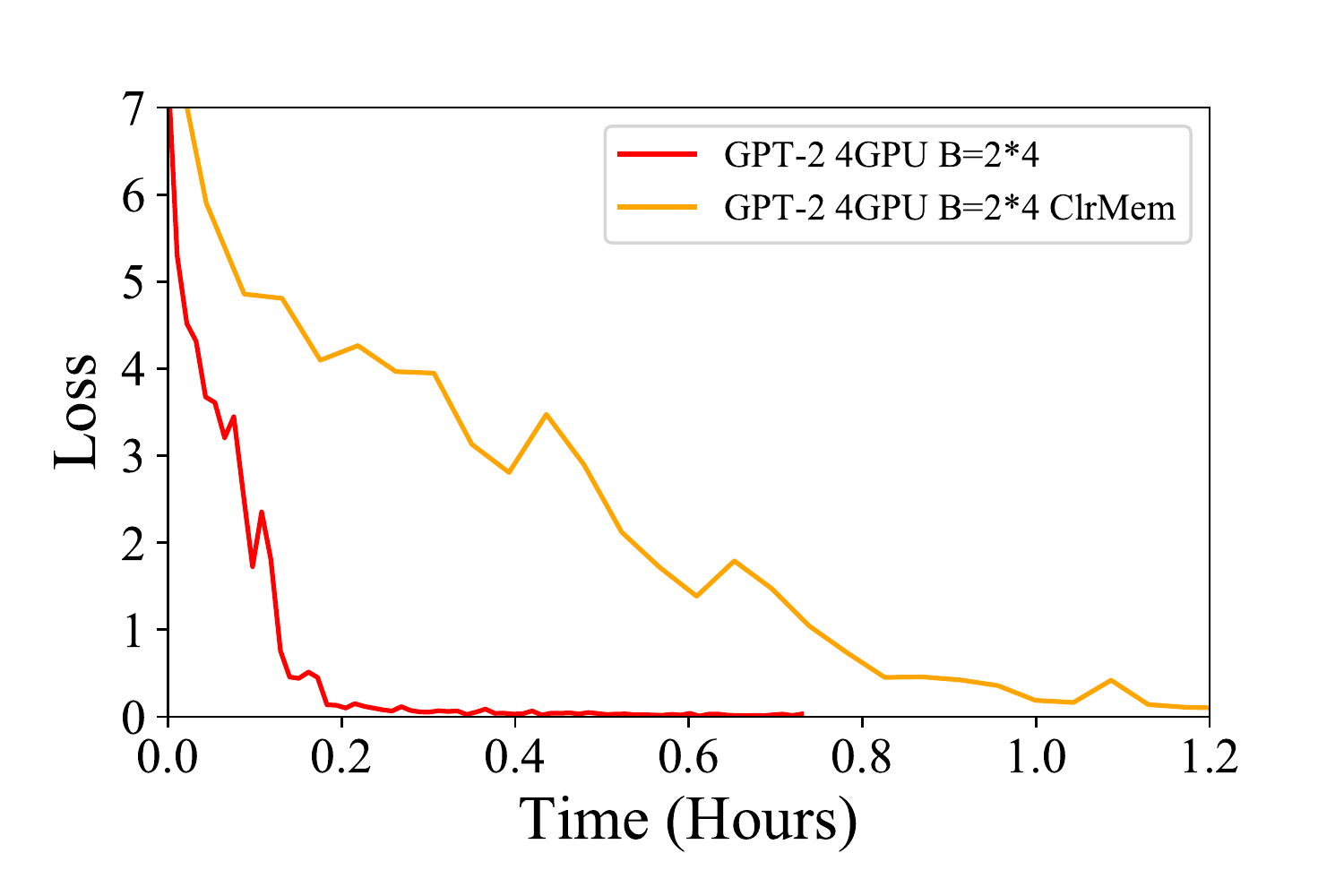}
            \caption[]%
            {{\small The loss-time curve of running GPT-2 under 4 GPUs with batch size 2*4 with memory clear and without memory clear.}}    
            \label{fig:dps-time-clrmem}
        \end{subfigure}
        \caption[]
        {\small The experimental results of the loss curve of GPT-2 under 4 GPUs before/after applying the memory clean code, with respect to number of training steps (a) and number of hours (b).}
        \label{fig:dps-time-step-clrmem}
    \end{figure*}

Compared to other strategies, this strategy is more engineering-based and has nothing to do with Formula \ref{equ:mem_calc}. If we represent the real memory usage as $\mathcal{M}_r$, and the marginal cost by the code itself as $\mathcal{M}_c$, we have $\mathcal{M}_r=\mathcal{M}+\mathcal{M}_c$. 

The marginal cost of unnecessary variables in research code is very common, as precise allocation and deallocation of data pointers in the Python code need a great command of the Python Language. A common problem is that few researchers notice to free the variables in each step and epoch. We take the code snippet below for an example.

\begin{lstlisting}[language={Python},numbers=left,numberstyle=\tiny,%frame=shadowbox,
   rulesepcolor=\color{red!20!green!20!blue!20},
   keywordstyle=\color{blue!70!black},
   commentstyle=\color{blue!90!},
   basicstyle=\ttfamily]
for epoch in range(epochs):
  for step in range(_len_ // batch_size):
    batch = ..
    out = model.forward()
    loss, logits = out[:2]
    loss.backward()
    optimizer.step()
\end{lstlisting}

The problem with this code is the variables allocated are still in the memory after each step ends. Note that these variables are in the CUDA memory instead of CPU memory because it would take too long for data to be transferred back and forth between GPU and CPU. Thus, the allocated CUDA memory cannot be freed. After each step, the GPU program not only recalculates the variables and replaces the original one, but also caches the original value in case of future use, leading to a serious memory expansion. 

To solve this problem, an examination of what variables are safe to be deleted should be conducted, and then the researcher should delete the structures and free the GPU cache. After that, it's recommended to use \texttt{del} followed by the \texttt{empty\_cache()} command. For example, the correct fix for the code above is adding these lines at the end of each step:

\begin{lstlisting}[language={Python},numbers=left,numberstyle=\tiny,%frame=shadowbox,
   rulesepcolor=\color{red!20!green!20!blue!20},
   keywordstyle=\color{blue!70!black},
   commentstyle=\color{blue!90!},
   basicstyle=\ttfamily]
# epoch:
  # step:
    del out, loss, logits
    torch.cuda.empty_cache()
\end{lstlisting}

The \texttt{empty\_cache()} command explicitly tells the CUDA compiler (\texttt{nvcc}) to clear the cache in memory. This command cannot be used without deleting the variable, because the run-time program is not able to find what cache to clear - both lines should be inserted at the same time.

As shown in Figure \ref{fig:dps-step-clrmem}, this approach has a drawback: the loss-time curve loses non-trivial performance for multi-GPU training. From the comparison of the experimental results, we see that clearing the memory in multi-GPU training takes lots of time. This is because flushing the cache will force the program to call a cleaner handler each time.

Conclusively, it's still recommended to use this strategy only for small models. For a more detailed explanation to multiple-GPU programming, you can refer to Section \ref{sec:data-parallel} and Section \ref{sec:model-parallel}.

\subsection{Change to CPU Memory} \label{sec:cpu-memory}

Changing the training process from GPU to CPU is another choice. As mentioned in \ref{sec:cuda-mem}, the general memory is much larger than the GPU memory, so it's harder to run out of memory. However, training on the CPU is extremely slow, so we recommend only placing a small bunch of memory-consuming computations on the CPU.

\subsection{Minimize Model}

Minimizing the model is one of the most difficult strategies for reducing $\mathcal{M}$. It's hard to conduct because the model varies from one to another, leading to uncertainty \cite{Laosong_2021}. In fact, this approach has even branched a subject itself, called model compression \cite{bucilua2006model}.

The basic strategies include changing two-level LSTM to one-level, replacing LSTM with GRU, reducing the number of convolution cores, and using as few \texttt{Linear} layers as possible \cite{Chenyun_2018}.

\subsection{Multi-GPU Data Parallel} \label{sec:data-parallel}

Data parallelization focuses on using multiple GPU workers to distribute the data to reduce $\mathcal{M}$. If we represent the number of GPUs working together as $k$, the memory usage $\mathcal{M}_i$ for each GPU worker $i$ now becomes the formula below.

\begin{align} \label{equ:mem_calc_data_parallel}
	\mathcal{M}_i &= p_m \cdot n + \frac{b \cdot p_o}{k} + \frac{p_b}{k}
\end{align}

As we see, for output-dense training (large $p_o$), this multi-GPU approach decreases $\mathcal{M}$ by a factor of approximately $k$.\footnote{For those models with a relatively small $p_o$, the effect may be decreased but still worth a try.} However, also note that $\frac{b}{k}$ must be an integer larger than one. If the batch size is 4 for 1 GPU, and 8 GPUs are available, then at most 4 GPUs can be used to apportion batch size, and the left 4 will get no data.

Here is another guess of the OOM error observed in Section \ref{sec:graphical}. In the experiment on DPS with each GPU apportioned batch size of 4, we found that the training process ran into an OOM error, but the benchmark with 1 GPU apportioned batch size of 4 did not. 

According to Formula \ref{equ:mem_calc_data_parallel}, together with the condition that we have the batch size on each GPU the same as the benchmark (4), the deducted memory usage on each GPU should be as described in Formula \ref{equ:mem_calc_data_parallel_large_batch}, which is exactly the same as Formula \ref{equ:mem_calc}, indicating the same memory usage with the benchmark. 

\begin{align} \label{equ:mem_calc_data_parallel_large_batch}
	\mathcal{M}_i &= p_m \cdot n + \frac{b \cdot p_o}{4}\cdot 4 + \frac{p_b}{4} \cdot 4
\end{align}

Except for the reason that the AllReduce algorithm still requires a master GPU to communicate, we suspect that $p_o$ (sum of output data for each layer) is not averagely distributed, as described in Formula \ref{equ:mem_calc_data_parallel_large_batch_bias}. We also take consideration of the marginal memory cost due to the engineering issue discussed in Appendix \ref{sec:free-vars}.

\begin{align} \label{equ:mem_calc_data_parallel_large_batch_bias}
	\mathcal{M}_i' &= p_m \cdot n + \frac{b \cdot (p_o+\Delta p)}{4}\cdot 4 + \frac{p_b}{4} \cdot 4 + \mathcal{M}_e
\end{align}

In the formula, $\Delta p_i$ is the sum of the difference of the output data for each layer in the model on each node, and $\mathcal{M}_e$ is the memory usage caused by engineering work like no memory deallocation and cache deletion (which was illustrated in Appendix \ref{sec:free-vars}).

\subsection{Multi-GPU Model Parallel} \label{sec:model-parallel}
	
The second type of multi-GPU strategy is the model parallel strategy. Different from the data-parallel strategy, it stores different parts of a model in different GPUs, with similar approaches like dispatching some data to the CPU for parallelization, as illustrated in Section \ref{sec:cpu-memory}. It's pretty important for the researcher to figure out which part of the model causes the OOM issue and then apportion that part to another GPU for storage and computation.

\end{document}